\documentclass[fleqn,usenatbib]{mnras}

\usepackage[T1]{fontenc}
\usepackage{ae,aecompl}

\usepackage{graphicx}	% Including figure files
\usepackage{amsmath}	% Advanced maths commands
\usepackage{amssymb}	% Extra maths symbols
\usepackage{natbib}
\usepackage[usenames]{color}
\usepackage{tabularx}
\usepackage{hhline}
\usepackage{array}
\usepackage{subfig}
\usepackage{url}
\usepackage[usenames]{color}
\usepackage{multirow}
\usepackage{booktabs}
\usepackage{hyperref}
\usepackage{xspace}

\newcommand{\cref}[1]{{{#1}}}
\newcommand\srm{\scriptscriptstyle\rm}

\def\OIII{[O\,${\srm III}$]\,}
\def\OIIId{[O\,${\srm III}$]\,$\lambda\lambda$4959, 5007}

\def\hbeta{H\,$\beta$}
\def\halpha{H\,$\alpha$}

\def\OII{[O\,${\srm II}$]}

\def\HEofor{HE\,0435$-$1223}
\def\HOLI{H0LiCOW\xspace}
\def\WFItwenty{WFI\,2033$-$4723}
\def\HEeleven{HE\,1104$-$1805}
\def\kms{km\,s$^{-1}$}
\def\kmsMpc{\rm km\,s^{-1}\,Mpc^{-1}}

\title[Environment of \HEofor]{H0LiCOW\,II. Spectroscopic survey and galaxy-group identification of the strong gravitational lens system \HEofor}

\author[D. Sluse et al.] 
{D.~Sluse$^{1}$\thanks{dsluse@ulg.ac.be},
	A.~Sonnenfeld$^{2,3,4}$, 
	N.~Rumbaugh$^{5}$,
	C.~E.~Rusu$^{5}$,
	C.~D.~Fassnacht$^{5}$,
	\newauthor
	T.~Treu$^{3}$,
S.~H.~Suyu$^{6,7,8}$,
K.~C.~Wong$^{7,9}$,
M.~W.~Auger$^{10}$,
V.~Bonvin$^{11}$,
T.~Collett$^{12}$,
\newauthor
F.~Courbin$^{11}$,
S.~Hilbert$^{13, 14}$,
L.~V.~E.~Koopmans$^{15}$,
P.~J.~Marshall$^{16}$,
G.~Meylan$^{11}$,
\newauthor 
C.~Spiniello$^{6}$,
M.~Tewes$^{17}$
\\
$^{1}$ STAR Institute, Quartier Agora - All\'ee du six Ao\^ut, 19c B-4000 Li\`ege, Belgium\\
$^{2}$ Physics Department, University of California, Santa Barbara, CA, 93106, USA \\
$^{3}$ Department of Physics and Astronomy, University of California, Los Angeles, CA 90095, USA \\
$^{4}$ Kavli IPMU (WPI), UTIAS, The University of Tokyo, Kashiwa, Chiba 277-8583, Japan \\
$^{5}$ Department of Physics, University of California, Davis, CA 95616, USA\\
$^{6}$ Max Planck Institute for Astrophysics, Karl-Schwarzschild-Strasse 1, D-85740 Garching, Germany\\
$^{7}$ Institute of Astronomy and Astrophysics, Academia Sinica, P.O.~Box 23-141, Taipei 10617, Taiwan\\
$^{8}$Physik-Department, Technische Universit\"at M\"unchen, James-Franck-Stra\ss{}e~1, 85748 Garching, Germany \\
$^{9}$ National Astronomical Observatory of Japan, 2-21-1 Osawa, Mitaka, Tokyo 181-8588, Japan\\
$^{10}$ Institute of Astronomy, University of Cambridge, Madingley Road, Cambridge CB3 0HA, UK \\
$^{11}$ Laboratoire d'Astrophysique, Ecole Polytechnique F{\'e}d{\'e}rale de Lausanne (EPFL), Observatoire de Sauverny, CH-1290 Versoix, Switzerland\\
$^{12}$ Institute of Cosmology and Gravitation, University of Portsmouth, Burnaby Rd, Portsmouth PO1 3FX, UK \\
$^{13}$ Exzellenzcluster Universe, Boltzmannstr. 2, 85748 Garching, Germany\\
$^{14}$ Ludwig-Maximilians-Universit{\"a}t, Universit{\"a}ts-Sternwarte, Scheinerstr. 1, 81679 M{\"u}nchen, Germany \\
$^{15}$ Kapteyn Astronomical Institute, University of Groningen, PO Box 800, NL-9700 AV Groningen, The Netherlands\\
$^{16}$ Kavli Institute for Particle Astrophysics and Cosmology, Stanford University, 452 Lomita Mall, Stanford, CA 94035, USA\\
$^{17}$ Argelander-Institut f\"ur Astronomie, Auf dem H\"ugel 71, D-53121 Bonn, Germany
}

\date{Accepted XXX. Received YYY; in original form ZZZ}

\pubyear{2016}

\begin{document}
\label{firstpage}
\pagerange{\pageref{firstpage}--\pageref{lastpage}}
\maketitle

% Abstract of the paper
\begin{abstract}
	
  {Galaxies located in the environment or on the line of sight towards gravitational lenses can significantly affect lensing observables, and can lead to systematic errors on the measurement of $H_0$ from the time-delay technique. We present the results of a systematic spectroscopic identification of the galaxies in the field of view of the lensed quasar \HEofor\, using the W.~M.~Keck, Gemini and ESO-Very Large telescopes. Our new catalog triples the number of known galaxy redshifts in the direct vicinity of the lens, expanding to 102 the number of measured redshifts for galaxies separated by less than 3\arcmin\,from the lens. We complement our catalog with literature data to gather redshifts up to 15\arcmin\, from the lens, and search for galaxy \cref{groups or clusters} projected towards \HEofor. We confirm that the lens is a member of a small group that includes at least 12 galaxies, and find 8 other group candidates near the line of sight of the lens. The {\it flexion shift}, namely the shift of lensed images produced by high order perturbation of the lens potential, is calculated for each galaxy/group and used to identify which objects produce the largest perturbation of the lens potential. This analysis demonstrates that i) at most three of the five brightest galaxies projected within 12\arcsec\,of the lens need to be explicitly used in the lens models, and ii) the groups can be treated in the lens model as an external tidal field (shear) contribution.}
    %The statistical impact of the groups and voids on the lens model is presented in a companion paper \HOLI III. The exhaustive lens modeling of \HEofor\, used for cosmological inference, including all the environmental sources of systematic errors, is presented in another companion paper \HOLI IV. We make our spectroscopic catalog and reduced spectra publicly available. } 

%These ingredients of the lens models are 
\end{abstract}

% Select between one and six entries from the list of approved keywords.
% Don't make up new ones.

%%%%%%%%%%%%%%%%%%%%%%%%%%%%%%%%%%%%%%%%%%%%%%%%%%

%%%%%%%%%%%%%%%%% BODY OF PAPER %%%%%%%%%%%%%%%%%%

%-------------------------------------------------------------------------------

\section{Introduction} 
\label{sec:intro}

Ongoing and upcoming cosmological studies deeply rely on the accurate knowledge of the Hubble constant, $H_0$ \citep{Hu2005, Suyu2012, Weinberg2013}. The measurement of $H_0$ has long been controversial \citep[e.g.][]{KOC02, KochanekSchechter2004}, but in the past decade several techniques have  measured $H_0$ with a relative uncertainty much smaller than 10\% \citep{Freedman2010, Humphreys2013, Suyu2013a, Riess2016}. In order to reach the goal of the next decade of cosmological experiments, and be able to e.g. unveil the nature of dark energy, it is necessary to pin down the accuracy on $H_0$ at the percent level. This is an ambitious goal and in order to identify unknown systematic errors, it is mandatory to gather several independent constraints on $H_0$ \citep{Weinberg2013, Riess2016}. The gravitational time-delay technique \citep{Refsdal1964}, applied to a large number of lensed systems, is one of the few techniques allowing one to reach percent precision on $H_0$ \citep{Suyu2012}. Among the various cosmological probes, it is also the most sensitive to $H_0$ \citep[e.g.][]{Jackson2007, Freedman2010}. By measuring the time delay $\Delta t$ between pairs of lensed images, and modeling the mass distribution of the lens galaxy, the time delay distance $D_{\Delta t}$ can be inferred. As summarized in a recent review by \cite{Treu2016}, the technique has long been plagued by poor time-delay measurements, invalid assumptions about the lens mass profile and systematic errors. However, times have changed. It has been demonstrated that an exhaustive study of a lensed quasar with high quality lightcurves \citep[B1608+656;][]{Fassnacht2002} allows the measurement of $H_0$ for a single system with a precision of 6\% \citep{Suyu2010}. In addition, it was shown that the time-delay technique leads to tight constraints on the other cosmological parameters comparable to those from contemporary Baryon Acoustic Peak studies, when each probe is combined with the Cosmic Microwave Background \citep{PLXVI, Anderson2014, Planck2015XIII, Ross2015}. 

The improved precision of the time delay technique stems from the combination of several ingredients. First, \cref{the COSmological MOnitoring of GRAvItational Lenses (COSMOGRAIL)} has been running for over a decade, gathering exquisite high cadence photometric data for tens of lensed quasars \citep{Eigenbrod2005, Tewes2013}. Those unprecedented high quality lightcurves combined with new curve shifting algorithms \citep{Tewes2013a} now enable time-delay measurements down to a few percent accuracy \citep{Bonvin2016, Liao2016}. Second, advanced modeling techniques that use the full surface brightness of the multiple lensed images, containing thousands of pixels as data points, are now used to constrain the lens mass distribution \citep{Suyu2009}. Third, independent constraints on the lens potential, obtained from the measurement of the lens velocity dispersion \citep{Romanowsky1999, Treu2002}, are now combined with the lens models, enabling one to reduce the impact of the mass-sheet degeneracy\footnote{The impact on cosmographic inference of other degeneracies among lens models, such as the source position transformation \citep{SS14, Unruh2016}, that does not leave the time-delay ratio invariant, still needs to be quantified.} \citep{FGS85, SS13} on the lens models. Finally, the direct lens environment and the line-of-sight galaxies are studied in detail \citep{Keeton2004,Fassnacht2006}. The observed galaxy counts in the vicinity of the lens are compared to galaxy counts from ray tracing through cosmological simulations to derive a probability distribution of the external convergence $\kappa_{\rm ext}$ produced by over- and under-densities along the line of sight \citep{Hilbert2007, Fassnacht2011}. 

The \HOLI project ($H_0$ Lenses in COSMOGRAIL's Wellspring) aims at achieving better than 3.5\% accuracy on $H_0$. To reach this goal, we have gathered a sample of five lenses (B\,1608$+$656, RX\,J1131$-$1231, \HEofor, \HEeleven, \WFItwenty) for which we apply our modeling technique on archival and Cycle 20 \emph{HST} data (PI Suyu). The project, together with cosmographic forecasts based on the full sample, is presented in  \HOLI Paper I (Suyu et al., submitted). The first two systems have been analyzed \citep{Suyu2010, Suyu2013a}. To tackle systematic errors in the other three systems, a stellar velocity dispersion for the lenses and a study of the lens environments are needed. In this paper, we focus on the spectroscopic identification of the brightest galaxies in the field of view of \HEofor, a quadruply imaged quasar at $z_{\rm s} = 1.693 \pm 0.001$ lensed by a foreground elliptical galaxy at $z_{\rm d} = 0.4546 \pm 0.0002$ \citep{Wisotzki2002, Morgan2005, Sluse2012b}. The main objective of this work is to measure the spectroscopic redshifts of most of the bright galaxies in the central region around \HEofor\, (i.e. about 100 galaxies), a necessary observable to measure the contribution of individual galaxy halos to the surface mass density projected towards \HEofor\, \citep{Hilbert2007, Hilbert2009, Greene2013, Collet2013}. Our secondary objective is to identify major groups and/or galaxy cluster(s), as well as individual galaxies, at the redshift of the main lens but also along the line of sight, that would perturb non linearly the gravitational potential of the main lensing galaxy. For that purpose we complement our data with the spectroscopic catalog compiled by Momcheva et al. (\citeyear[][hereafter MOM15]{Momcheva2015}) that gathers redshifts of $\sim$ 400 galaxies (about 30 galaxies are duplicated with our catalog) over a 30\arcmin$\times$30\arcmin\, field centered on \HEofor. The spectroscopic redshift measurements are an important ingredient of the statistical analysis of the line of sight towards \HEofor\, carried out in the companion \HOLI Paper III (Rusu et al., submitted). This companion paper presents a weighted count analysis of the galaxies in the field of view of \HEofor\, that is compared to galaxy counts from the Canada-France-Hawaii-Telescope Legacy Survey \citep[CFHTLenS, ][]{Heymans2012} and to galaxy counts from Millennium Simulation \citep{Springel2005, Hilbert2007, Hilbert2009}. This yields a probability distribution of convergence $\kappa_{\rm ext}$ produced by the other galaxies in the field. On the other hand, the redshifts of the galaxies closest in projection to the lens are included explicitly in the multi-plane lens modeling analysis of \HEofor\, presented in \HOLI Paper IV (Wong et al., submitted). Finally, Paper V (Bonvin et al., submitted) presents the time-delay measurements of \HEofor\,and the joint cosmographic inference from the three lensed systems analyzed to-date in \HOLI.

The paper is structured as follows. We present an overview of the data sets used, of the data reduction process and redshift measurements in Sect.~\ref{sec:data}. The methodology used to identify galaxy groups is explained in Sect.~\ref{sec:groups}. The galaxy groups identified with our algorithm and the spectra of the galaxies that \cref{are most likely to produce large gravitational potential perturbations} are presented in Sect.~\ref{sec:environment}. Section ~\ref{sec:model} quantifies the impact of individual galaxies and galaxy groups on the model. We use the \emph{flexion shift} to flag the systems that require explicit inclusion in the multi-plane lens models presented in \HOLI Paper IV. Finally, Sect.~\ref{sec:conclude} summarizes our main results. In this work, with the exception of the target selection that was based on $R-$band magnitude in the Vega system, photometric information comes from the deep multicolor imaging presented in \HOLI Paper III and uses the AB photometric system. 
For convenience, group radii and masses reported in this work assume a flat $\Lambda$CDM cosmology with cosmological parameters from \citep{Planck2015XIII}, namely  $H_0=67.7 \,\kmsMpc$, $\Omega_m$ =\,0.307. We stress that this choice has no impact on the group identification as the latter does not depend on a specific choice of cosmological parameters.

%-------------------------------------------------------------------------------

\section{Data} 
\label{sec:data}

Our data set combines multi-object spectroscopy obtained at Gemini-South, Keck, and ESO-Paranal observatories. We describe in Sect.~\ref{subsec:target} our target selection methodology. The observational setup, and data reduction techniques are described in Sect.~\ref{subsec:obs} \& ~\ref{subsec:reduc}. Finally, Sect. ~\ref{subsec:redshift} \& ~\ref{subsec:completeness} detail how the spectroscopic redshifts are measured, and evaluate the spectroscopic redshift completeness of our galaxy sample. The catalog and reduced spectra are available in electronic form \cref{at the  Centre de Donn\'ees astronomiques de Strasbourg (CDS) and from the \HOLI website\footnote{\url{www.h0licow.org/}}. The catalog contains 534 unique objects, including 368 redshifts exclusively reported by MOM15. Our new measurements expands to 169 the number of targeted objects separated by less than 3 arcmin from the lens. In that range, the new catalog contains 103 galaxies (but 16 have only tentative redshifts, and one is the lens), 42 stars, and 24 objects whose type could not be unambiguously determined and therefore lack redshift.} The first five entries of the full catalog are shown in Table~\ref{tab:catalog}.

% #counts using python on table 'cat0435_v1c_wdits.fits' / 'final/catHE0435_phot_spec_30arcmin_v1c.fits'
% ZQF = 0 and z > 0.01: 56  # secure gal. from us
% ZQF = 1 and z > 0.01: 16  # Unsecure gal. from us
% ZQF = 2 and z > 0.01: 24  (same if no constraint on z) # unknown
% ZQF > 2 and z > 0.01: 31  # gal from MOM15 (5 + lens have zQF=6)
% ZQF = 0 and z < 0.01: 37 # secure star
% ZQF = 1 and z < 0.01: 5  # unsecure star
% => # of secure galax: 56+31 = 87
% # of galax w. tentative z= 16
% # of stars: 37+5 = 42
% # of unknown: 24
% Checksum: 87+16+42+24 = 169 (including the lens)
%# of galax: 87+16 = 103-1 (lens) = 102 
%# of stars: 42

% Table generated direcly 
% tdfs[0:5].to_latex(buf='tabtest.tex', columns=['ID_1', 'OBJID', 'RA_1', 'DEC_1', 'z', 'z_err', 'zQF', 'Type'], index=False)
\begin{table*}
\caption{First lines of our spectroscopic redshift catalog. Columns \#1 to \#6 are objects name, ID, position (RA-DEC), redshift $z$ and its uncertainty $\sigma_z$. The last two columns display a quality flag and the object type. The full table is available in electronic form. \label{tab:catalog}}
\begin{tabular}{lccccccl}
	\toprule
	Name$^{1}$ &   ID &       RA &      DEC &         $z$ &     $\sigma_z$ &  zQF$^{2}$ &       Type$^{3}$ \\
	\midrule
	Momcheva\_201508\_obj10154 &  	10154 &  69.561980 & -12.287390 &  0.454600 &  0.000200 &  6 &  Unknown   \\
	FORS\_20131026\_obj0695    &      695 &  69.561350 & -12.288390 &  0.781733 &  0.000514 &  0 &  ETG-Sx      \\
	FORS\_20140124\_obj0357    &      357 &  69.562690 & -12.289600 &  0.780216 &  0.000531 &  0 &  ETG-Sx       \\
	FORS\_20140124\_obj0133    &      133 &  69.559851 & -12.286231 &  0.418679 &  0.000521 &  0 &  Starburst \\
	FORS\_20140124\_obj0188    &      188 &  69.560284 & -12.285444 &  0.456444 &  0.000536 &  0 &  ETG-Sx       \\
	\bottomrule
\end{tabular}
\begin{flushleft}
{\small {\bf {Notes:}} $(1)$ Format: Instrument\_date\_objID, where instrument is \emph{FORS}, \emph{Gemini} or \emph{Keck} if the redshift is derived from our survey, and Momcheva if the redshift comes from MOM15. The ``date'' in format yyyymmdd is the date of observation, or \emph{201508} for objects from MOM15. \\
$(2)$ The quality flags zQF=0/1/2 if the redshift is extracted from this program and 3,4,5,6 refer to objects from MOM15. zQF=0 for secure redshift; zQF=1 for tentative redshift; zQF=2 for unreliable/unknown redshift; zQF=3 for data obtained with LDSS-3; zQF=4 for data obtained with IMACS; zQF=5 for data obtained with Hectospec; zQF=6 for NED objects. \\
$(3)$ Type=ETG-Sx if CaK-H and/or G-band are detected; Type=Starburst if clear emission lines are observed, Type=M-dwarf for a M-dwarf star; Type=Star for other stellar-types; Type=Unknown if no identification could be done or if the spectrum is from an external catalog.}
\end{flushleft}
\end{table*}

\subsection{Target selection}
\label{subsec:target}

The targets were selected based on a $R$-band photometric catalog constructed using SExtractor \citep{Bertin1996} applied on archive images obtained with the FOcal Reducer and low dispersion Spectrograph for the Very Large Telescope (FORS1 and FORS2 at VLT). Because of the unavailability of deep frames obtained under photometric conditions, we had to construct an approximate photometric catalog from shallow $R$-band FORS1 images for which a photometric zero-point was available, and from deep FORS2 $z$-band frames lacking photometric calibration. By matching objects found in both catalogs, hence implicitly assuming a constant color term, we could get an approximate photometry of targets down to $R \sim 23.5$\,mag. Comparison of the photometry of the brightest objects in the field with SDSS-DR9 and USNO photometry suggested a photometric accuracy of $\sim 0.15$\,mag. This has been confirmed a-posteriori using the deep Subaru/Suprime-Cam $r-$band photometry presented in \HOLI Paper III. The photometry of the two catalogs agree with each other with a scatter on the difference of 0.17\,mag. In the analysis presented here we do not use that preliminary photometry but the most accurate one presented in  \HOLI Paper III. Keeping in mind the importance of identifying all the faint galaxies in the close vicinity of the lens, we have prioritized the spectroscopic targets using the following scheme. Any potential galaxy (i.e. objects with SExtractor flag \texttt{CLASS\_STAR} $<$ 0.98) with $R<23.5$\,mag located within a 30$\arcsec$ radius from the main lens was given highest priority (i.e. P1). Any potential galaxy with $R<21$\,mag located within 3$\arcmin$ from the lens was also flagged as high priority (P1). This selection towards \cref{bright objects was set to avoid missing the identification of of massive nearby galaxy clusters}. Medium priority (P2) objects were galaxies with $21 < R < 22.4$\, mag located in an annulus $0\arcmin.5 < r < 1 \arcmin$ from the lens. Finally, lower priority objects (P3) were those galaxies beyond 1$\arcmin$ from the lens (but within  $3\arcmin$), with $21 < R < 22$\,mag. \cref{Any object not entering in the above categories was used as a filler and targeted if free slits were available.} When possible, we tried to observe again the faintest targets (i.e. $R<22.4\,$mag) to increase the signal-to-noise ratio in their spectra. \cref{We have compared, a posteriori, our original object selection with the one we would have carried out based on the deeper Subaru/Suprime-Cam photometry (which has a magnitude limit of $r=25.94\pm 0.28$\,mag). We found that 1 (P1), 3 (P2) and 4 (P3) objects were missed in the original catalog. This corresponds to typically 10\% of missed targets. Those mismatches were caused by differences in SeXtractor parameters yielding inaccurate deblending rather than by the photometric inaccuracy of the original catalog. The impact of spectroscopic incompleteness on our analysis is discussed in Sect.~\ref{subsec:completeness} \& \ref{sec:environment}.} Figure~\ref{fig:field} shows the field around \HEofor\, targeted by our program. Targets with secured redshifts, tentative redshifts, failed redshift measurements, and unobserved galaxies, are respectively depicted with colored circles, colored boxes, black boxes and gray circles.

% Regarding "completeness" and selection bias: MXU_environment.figphotcompleteness()
\subsection{Observations}
\label{subsec:obs}

The largest data set has been obtained with the FORS2 instrument \citep{Appenzeller1998} mounted at the Cassegrain focus of the UT1 (Antu) telescope (PID: 091.A-0642(A), PI: D.\,Sluse). The instrument was used in its multi-object spectroscopy mode with exchangeable masks (MXU), where masks are laser cut at the location of the targets. The GRIS300V grism\,+\,GG435 blocking filter were used to ensure a large spectral coverage (see Table~\ref{tab:data}) in order to maximize the range of redshift detectability. Four masks with different orientations on the sky were employed to best cover the 6\arcmin\,$\times$\,6\arcmin\, field of view centered on \HEofor. Each mask was composed of approximately 40 slits of 1\arcsec\, width and typically 8\,\arcsec\,long (the slit length was reduced by a few arcseconds for some objects to avoid overlap of spectra). This slit length was sufficiently large compared to the seeing and typical target size to allow the use of regions of a few arcseconds around the object to carry-out adequate sky subtraction. In addition, owing to the spatial sampling of 0\farcs25/pixel, we sometimes included 2 nearby objects in the same slit to maximize the number of observed targets. Observations were obtained under seeing condition generally better than $0\farcs8$\,FWHM ($R-$band) at airmasses ranging between 1.024 $< \sec(z) < $ 1.519. FORS2 data were obtained in service mode between October 2013 and January 2014 (i.e. 2013-10-26, 2014-01-24, 2014-01-27). 

Another ensemble of 51 spectra were obtained with the Gemini Multi-Object Spectrographs \citep[GMOS;][]{Hook2004} at the Gemini-South telescope, used in multi-object spectroscopy mode (PID: GS0213B-Q-28, PI: T. Treu). The observing strategy was the same as for FORS data. The additional masks provided an increase in the completeness in the vicinity of the lens where most of the highest priority targets are located. We used the R400 grating with GG455 filter for our observations, providing a wavelength coverage of most of the visible spectrum with a resolving power of 1100.  Each target was observed through a 6\arcsec$\times$1\arcsec\, slitlet. Three slitlet masks, covering a 2.3\arcmin\,$\times$2.3\arcmin\, field of view centered on the lens, were used to observe all the targets. Dithering in both spatial and spectral direction \cref{(i.e. changing the central wavelength of the grating by 10\,\AA)} was applied between exposures to reduce the impact of bad pixels. Observations were carried out in service mode on the nights 2013-11-22 and 2013-11-23.

% 32 targets if we include m4 masks but I have no redshift measurement associated to mask #4. 
Spectra of 26 targets were obtained in 2008 and 2011 using the Keck Low Resolution Imaging Spectrometer \citep[LRIS; ][]{Oke1995} instrument (PI: Fassnacht). This spectrograph divides the beam into a red and blue arms whose light is dispersed with independent sets of grisms/gratings and collected by two different CCDs that can operate simultaneously. The first set of observations were taken on 2008 Nov 24 under moderate conditions with seeing varying between 1 and 2\arcsec. The dispersing elements were the 600/7500 grating on the red side, giving a dispersion of 1.28\,\AA\ pix$^{-1}$ and a central wavelength set to be roughly 6600\,\AA, and the 600/4000 grism on the blue side, giving a dispersion of 0.63\,\AA~pix$^{-1}$.  We obtained five exposures through one slitmask, each of 1800~s, interspersed with calibration observations of arclamps and internal flats. The second set of observations were obtained on 2011 Jan 05, where two slitmasks were observed. For these masks, the red-side dispersing element was the 831/8200 grating, with a dispersion of 0.58\,\AA pix$^{-1}$ and a central wavelength of roughly 6800\,\AA, while the 600/4000 grism was once again used on the blue side.  Each of these slitmasks was observed for 1200~s each. For all masks, a slit width of 0.7\arcsec\, was used and the D560 dichroic was used to split the incoming light between the red and blue arms.

\begin{table}
	\caption{{\bf Overview of the data set}. The columns list respectively the instrument used (LRIS-B and LRIS-R correspond to the blue and red arms of LRIS), the number of masks, the total number of spectra obtained, the approximate resolving power $R$ of the instrument at central wavelength, the typical wavelength range covered by the spectra (spectra do not always cover the full wavelength range depending of their exact object location in the field), and the exposure time per mask. Note that the \# of spectra includes duplicated objects.}
	\label{tab:data}
	\begin{tabular}{cccccc}
		\toprule
		Instrument & \# of  & \# of & $R$ & $\lambda_1-\lambda_2$  & Exp  \\
		 & Masks & spectra & & (\AA) & (s) \\ 
		\midrule
		FORS2  & 4 & 156 & 440 &4500-9200 & 2$\times$1330\\
		GMOS  & 3 & 51 & 1100 & 4400-8200& 4$\times$660 \\	
		LRIS-B  & 3 & 26 & 1200 & 3300-5400 & 5$\times$1800\\
		LRIS-R$^\dagger$   & 1 & 10 & 1700 & 5500-8000 & 5$\times$1800 \\
		LRIS-R$^\ddagger$  & 2 & 16 & 2300 & 5600-8000 & 6$\times$1200 \\
	    \bottomrule
	    
	\end{tabular}
{\small \\
	{\bf Notes:} $^\dagger$ data from 2008-11-24 ; $^\ddagger$ data from 2011-01-05
}
\end{table}

\subsection{Data reduction}
\label{subsec:reduc}

The FORS2 data have been reduced using the ESO \texttt{reflex} environment \citep{Freudling2013}. Version 2.2 of the FORS2 pipeline has been used, yielding wavelength and flux calibrated two-dimensional (2-D) spectra for each individual exposures. The reduction cascade, described extensively in the FORS pipeline user manual \citep{Izzo2013}, includes the standard MXU spectroscopic data reduction steps, namely bias and dark current subtraction, detection of the individual slits and construction of extraction mask, correction of the science frames with normalized flat-field, sky subtraction, wavelength calibration and geometric correction. Default parameters of reduction routines were used, except for the wavelength calibration where a polynomial of degree $n=4$ gave the best solution with residuals distributed around 0, a RMS of typically 0.1-0.2 pixels at all wavelengths and a model accuracy derived by matching the wavelength solution to the sky lines, to 0.2\,\AA. Cosmic rays have not been removed within the pipeline but separately, using the LA-COSMIC routine \citep{Vandokkum2001}. Extraction was subsequently performed using customized Python routine fitting 1-D Gaussian profile on each wavelength bin of the rectified 2-D spectrum. When multiple objects were present in the same slit, a sum of profiles centered on each target was used for the extraction. For each mask, a set of two exposures were obtained. The one-dimensional spectra extracted on individual exposures were finally co-added. 

GMOS data were reduced using the Gemini IRAF\footnote{IRAF is distributed by the National Optical Astronomy Observatories, which are operated by the Association of Universities for Research in Astronomy, Inc., under cooperative agreement with the National Science Foundation.} package. Dedicated routines from the {\it gemini-gmos} sub-package were used to perform bias subtraction, flat-fielding, slit identification, geometric correction, wavelength calibration and sky subtraction on each exposure, producing a wavelength calibrated 2-D spectrum for each slitlet. Wavelength calibration was done in interactive mode:  we visually inspected the automatic identification of arc lamp lines produced by the pipeline and applied corrections in cases of mis-identification. We then used a custom Python script to extract one dimensional (1-D) spectra for each detected object in each slitlet and to co-add spectra from different exposures of the same mask.

The Keck/LRIS data were reduced with a custom Python package that has been developed by our team. This package automatically performs the standard steps in spectroscopic data calibration including overscan subtraction, flat-field correction, rectification of the two-dimensional spectra, and wavelength calibration.  For the red-side spectra, the wavelength calibration was derived from the numerous night sky-lines in the spectra, while on the blue side the arclamp exposures were also used.  The 1-D spectra were extracted from each exposure through a given slitmask using gaussian-weighted profiles.  The extracted spectra were co-added using inverse-variance weighting.

\subsection{Redshift measurement}
\label{subsec:redshift}

\begin{figure*}
	% Final figure generated with MXU_environment.makefigFOV():
	\includegraphics[width=\linewidth]{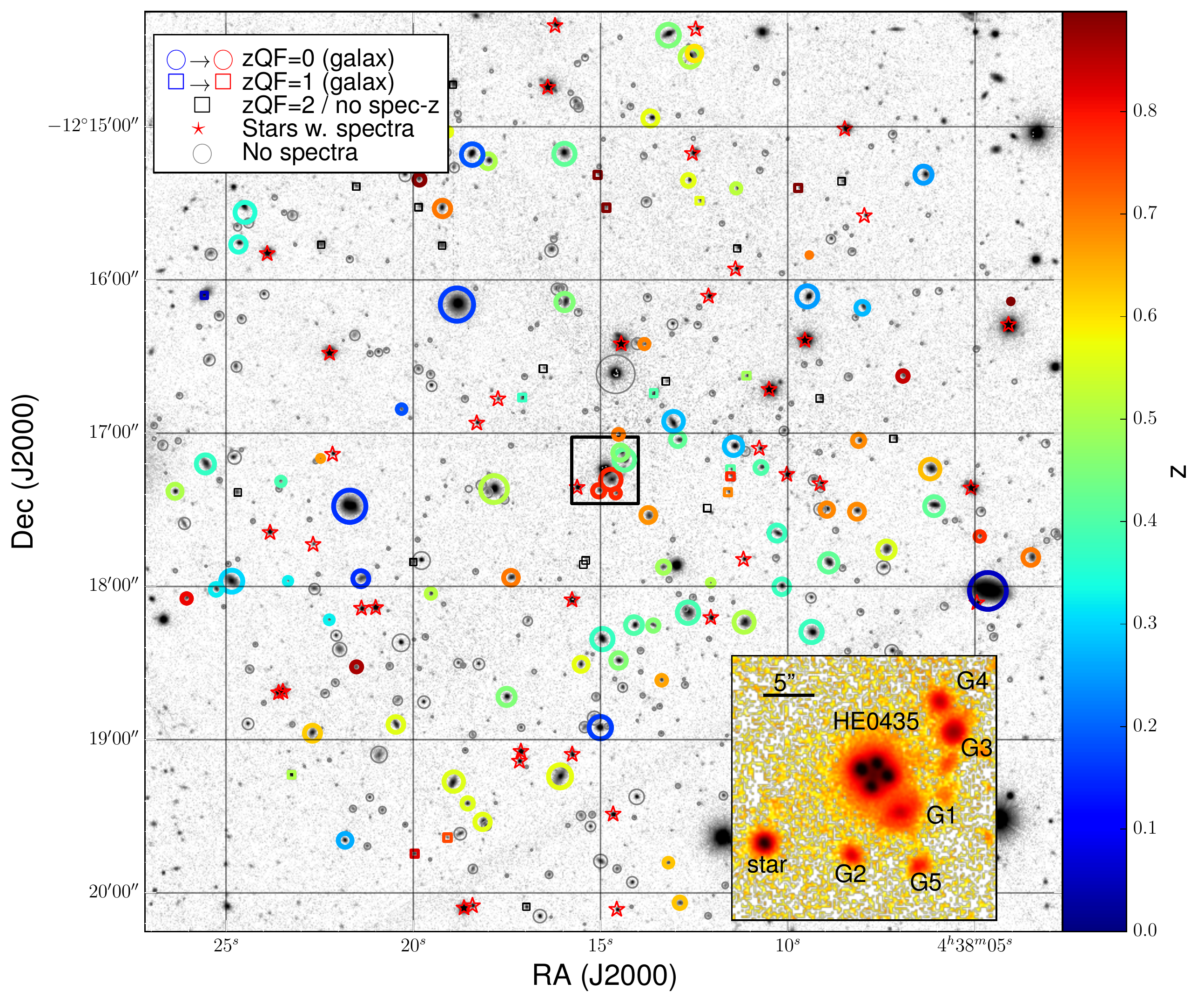}
	\caption{Overview of the spectroscopic redshift obtained from our new and literature data in a field of view of $\sim$ 3\arcmin\,$\times$3\arcmin\, around \HEofor\,(black box and inset panel). Spectroscopically identified stars are marked with a red "Star" symbol, while galaxies are marked with a circle whose size scales with its $i$-band magnitude (largest colored circle correspond to $i \sim $17\,mag, smallest to $i\sim $23\,mag), and color indicates the redshift (right color bar). A gray circle is used when no spectroscopic data are available. Galaxies that have been targeted but for which no spec-z could be retrieved are shown as open black squares, those with a tentative redshift (zQF = 1, see Table~\ref{tab:catalog}) with a colored square. The background frame shows the central region of deep 600\,s $i-$band image obtained with Subaru/Suprime-Cam and presented in \HOLI Paper III.}
	\label{fig:field}
\end{figure*}

The redshift measurements of the FORS2 \cref{(151 objects)}, GMOS \cref{(51 objects), and LRIS (26 objects)} data were performed by cross-correlating the 1-D spectra with a set of galactic \cref{(Elliptical, Sb, only galactic emission lines, quasar)} and stellar \cref{($G$, $O$, $M1$, $M8$, $A$ spectral types, all-stars)} templates using the \texttt{xcsao} task, part of the \texttt{rvsao} IRAF package (version 2.8.0). The package was used in interactive mode, excluding regions where the sky subtraction was not optimal. The redshift measurement was then flagged as secure \cref{(70\% of the measurements)}, tentative \cref{(15\% of the measurements)} or unsecure \cref{(15\% of the measurements)} based on the quality of the cross-correlation, signal-to-noise and number of emission/absorption lines detected. The formal uncertainty on the redshift from this procedure depends only on the width and peak of the cross-correlation. This formal uncertainty is smaller than the systematic error on the wavelength calibration. The latter has been derived by comparing redshifts of objects in common with the catalog\footnote{Only 8 galaxies have redshift measurements from both Gemini and FORS, and a handful from Keck and FORS, which limits our ability to perform internal comparisons.} published by MOM15 {\cref{(see Appendix~\ref{Appendix:redshift} and Fig.~\ref{fig:redshiftcompa})}}. The 30 galaxies in common with that catalog\footnote{We only consider objects with the same redshift and with flags 3 and 4, i.e. we exclude objects that are not new measurements from MOM15 but included in their catalog.} reveal a systematic offset $\delta z \sim$ -0.0004 between the two samples, or $\sim$ 1 pixel $\sim$ 3.3\,\AA\, in our wavelength calibration (i.e. about five times larger than the one derived along the reduction). This translates into a velocity offset $\delta v \sim 120$\,\kms. We account for this error in the following way. On one side, we subtract $\delta z \sim$ 0.0004 from the FORS redshifts, and on the other hand, we add quadratically an error $\sigma_z = 0.0005$ to the formal redshift error. This uncertainty has a negligible impact on our group detections compared to other sources of errors (see Sect.~\ref{subsec:groups}).

% NOTE THAT for WFI2033 ONLY 3 galax. from VLT are in common w. Momcheva 2015 ; 7 objects in Common between FORS and MUSe and deltaz ~ 5.57e-5+/-0.0001 => FORS and MUSE do not show systematic offset for WFI2033. 
% NOTE THAT FOR H1104, comparison with MOMCHEVA leads to 24 galaxies in common, and delta_z ~ zVLT-z_MOM = 0.00017+/- 0.000267 and mean error on redshift of 0.000127 => No discrepancy between MOMCHEVA and us. 

The comparison between multiple data sets also provides a good way to flag incorrect redshift measurements. Table~\ref{tab:badredshifts} lists the three objects that have been reported in MOM15 with a redshift significantly different from ours. Two of the redshift estimates from MOM15 are tentative literature measurements from \cite{Morgan2005}. The redshift of these galaxies, labeled G09 and G10 in \cite{Morgan2005}, was then based on a possible detection of \OII\, line. Our spectra, as well as \emph{HST} images, show that these objects are stars in our Galaxy. The third object (ID 11182 in MOM15) has a complex morphology and could potentially be a blend of two objects. We clearly detect \hbeta, \halpha, and \OIIId\, emission at a redshift $z=0.1537$. The redshift $z=0.5484$ proposed by MOM15 roughly matches a mis-identification of \OIII\,$\lambda\,5007$ as \OII\,$\lambda\,3727$ emission, which would explain the observed discrepancy. No groups are detected at the redshifts of those misidentified objects (Sect.~\ref{subsec:groups}). 

% Note that I remove object with ID 492 (Momcheva 9713) RA : 69.55255 ; Dec: -12.24246 ; because I think this is a different object. The two systems (mine and Momcheva) are located 2.4" from each other and this is larger than the average separation (0.4")

\begin{table*}
	\caption{Objects with significantly different redshifts in MOM15 and in our catalog. The last comment briefly summarizes the reason of the likely mis-identification in MOM15 (see Sect.~\ref{subsec:redshift} for more details). }
	\label{tab:badredshifts}
	\centering
	\begin{tabularx}{0.92\linewidth}{llllll}
		\toprule
		 (RA,DEC) & ID-MOM & ID & $z_{\rm MOM}$ ($\sigma_{z}$) & $z$ ($\sigma_z$) & Note \\
		\midrule
		
		(69.57627, -12.28224)	&10541	&251	&0.3380 (2.0E-4)	&0.	(0.001) & Based on spurious \OII. \\
		(69.57391, -12.27961)	&10425	&249	&0.3691 (2.0E-4)	&0.  (0.001)& Based on spurious \OII.   \\
		(69.58917, -12.29916)	&11182	&95	    &0.54839 (2.3E-4)	&0.15307 (9.3E-5) & Mis-identified \OII\,or blend of 2 objects. \\
%		(69.58012, -12.25887)	&10800  &1104   &0.70287 (3.0E-4)   &0.557   (0.001) & Unknown \\  % Wrong preliminary redshift
		% 1104 is currently removed from the Keck catalog
		\bottomrule
		% NOT a duplicate
		%69.55255	-12.24246	69.55204	-12.24204	9713	492	0.50345	3.0E-4	0.599028078285	1.32675118865E-4
		
	\end{tabularx}
	
\end{table*}

\subsection{Completeness of the spectroscopic redshifts}
\label{subsec:completeness}

% 356 objects are present up to 15 arcmin, but there is the lens that I remove, as well as 12 stars 
For the analysis presented in this paper, we have complemented our data with the spectroscopic catalog of MOM15 \cref{(343 new galaxies separated by up to 15\arcmin\, from the lens)}, and with $i-$band magnitudes (i.e. $i'$ filter from Subaru/Suprime-Cam, similar to SDSS-$i$ filter) from \HOLI Paper III. 

\begin{figure}
	% Figure generated with MXU_environment.makefigmag()
	\includegraphics[width=1.0\linewidth]{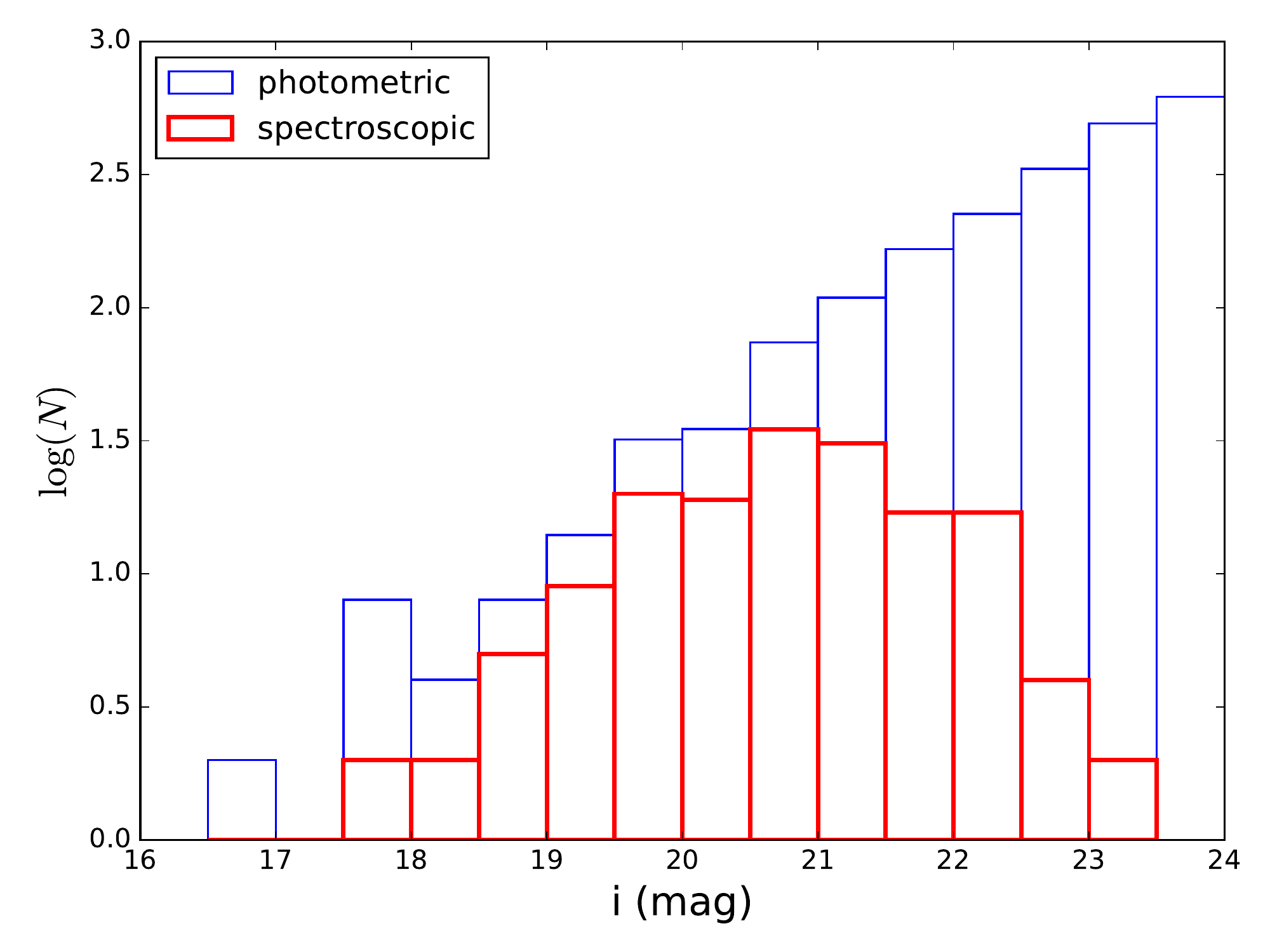}
	\caption{Apparent $i-$band magnitude histogram (log scale) of all the galaxies (thin blue) located within 6\arcmin\, of \HEofor\, and of the subsample of galaxies with a spectroscopic redshift (thick red). }
	\label{fig:magdistrib}
\end{figure}

\begin{figure}
	% Figure generated with MXU_environment.makefigcompleteness(cmulative=True)
	\includegraphics[width=1.0\linewidth]{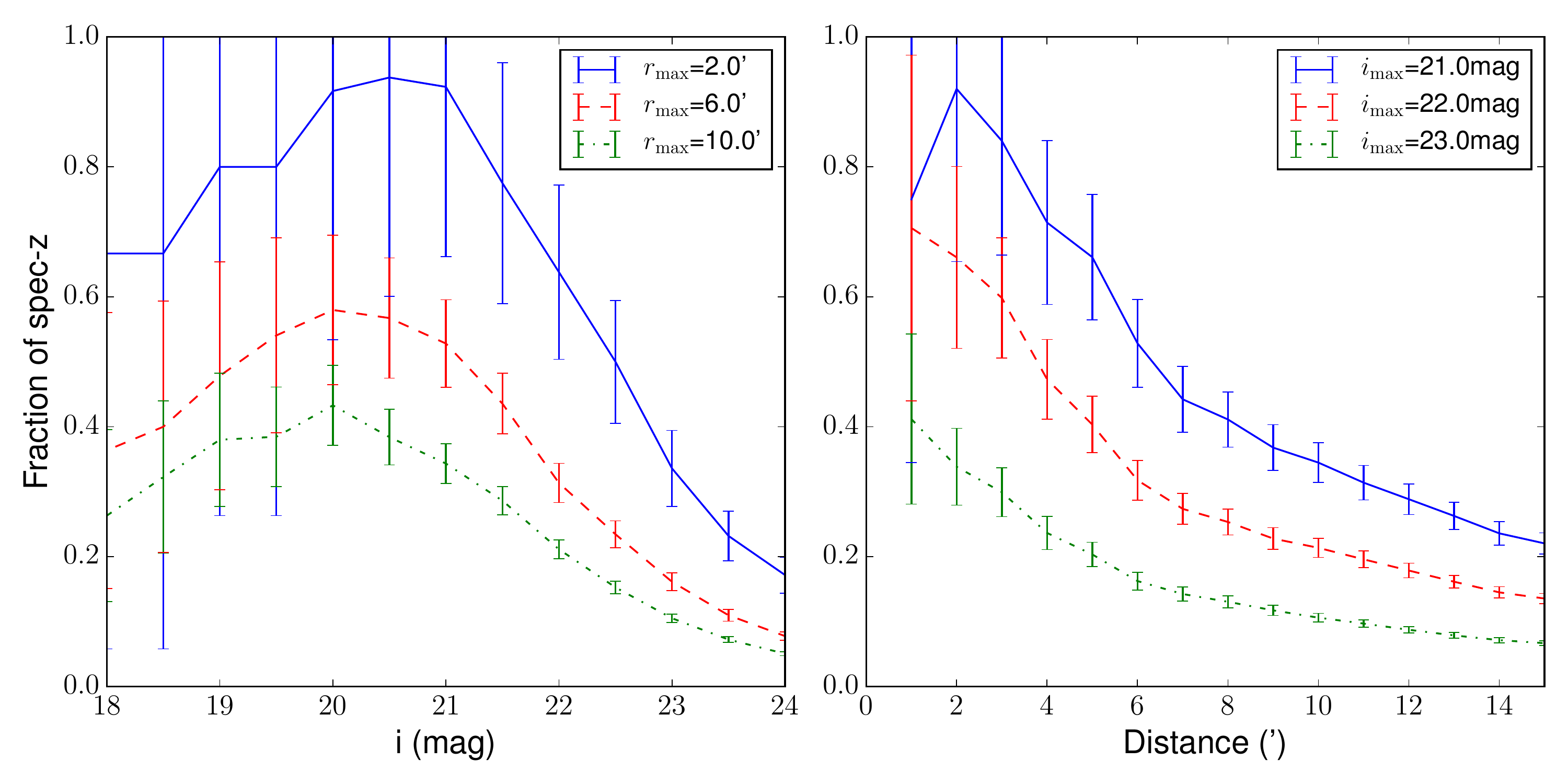}
	\caption{{\bf Left:} Fraction of spectroscopic redshifts (unsecure redshifts are not included) as a function of the maximum $i$-band magnitude of the sample, for three different \cref{radii 2\arcmin (solid-blue), 6\arcmin (dashed-red), 10\arcmin} (dashed-dotted-green). {\bf Right:} Fraction of spectroscopic redshifts as a function the maximum distance to the lens for three different limiting magnitude ($i_{\rm max} = 21$\,mag in solid-blue; $i_{\rm max} = 22$\,mag in dashed-red, $i_{\rm max} = 23$\,mag in dotted-dashed-green). The error bars are the Poisson noise derived from the number of objects studied. }
		
\label{fig:completeness}
\end{figure}

We evaluate the spectroscopic redshift completeness as a function of various criteria by comparing our spectroscopic and photometric catalogs. Figure~\ref{fig:magdistrib} shows, as a function of $i-$band magnitude, the number of galaxies (total, and with secure spectroscopic redshift, hereafter \emph{spec-z}) in the field of the lens. The number of galaxies with a secure spec-z drops significantly above $i=22.5$\,mag, as expected from our observational setup. Another important piece of information for our analysis is the completeness of our sample as a function of the magnitude of the galaxies and of the distance to the lens. Figure~\ref{fig:completeness} shows that our completeness is higher than 60\% in the inner 2\arcmin\, around the lens for galaxies brighter than $i\sim 22$\,mag. At larger distances, or fainter magnitude cutoff, the completeness of the spectroscopic catalog drops below 30\%. 

Identifying galaxy groups requires a high spectroscopic completeness over the chosen field of view. Based on Figure~\ref{fig:completeness}, we have decided to limit our search for groups to a maximum distance of $r\sim$\,6\arcmin\,of the lens. With this radius, we cover a region \cref{$\sim$ 3} virial radius $R_{\rm vir}$ of a typical group at $z=0.4\pm0.2$ (i.e. $R_{\rm vir} \sim 1$\,Mpc or $\theta_{\rm vir} \sim 2\arcmin$), and are complete at more than 50\% down to $i=22$\,mag. We show in Sect.~\ref{sec:model} that this is sufficient to identify groups that produce high order perturbations of the gravitational potential of \HEofor.  

\begin{figure}
	% Figure generated with MXU_environment.makefigcompleteness2(cumulative=False, rmax=360)
	\includegraphics[width=1.0\linewidth]{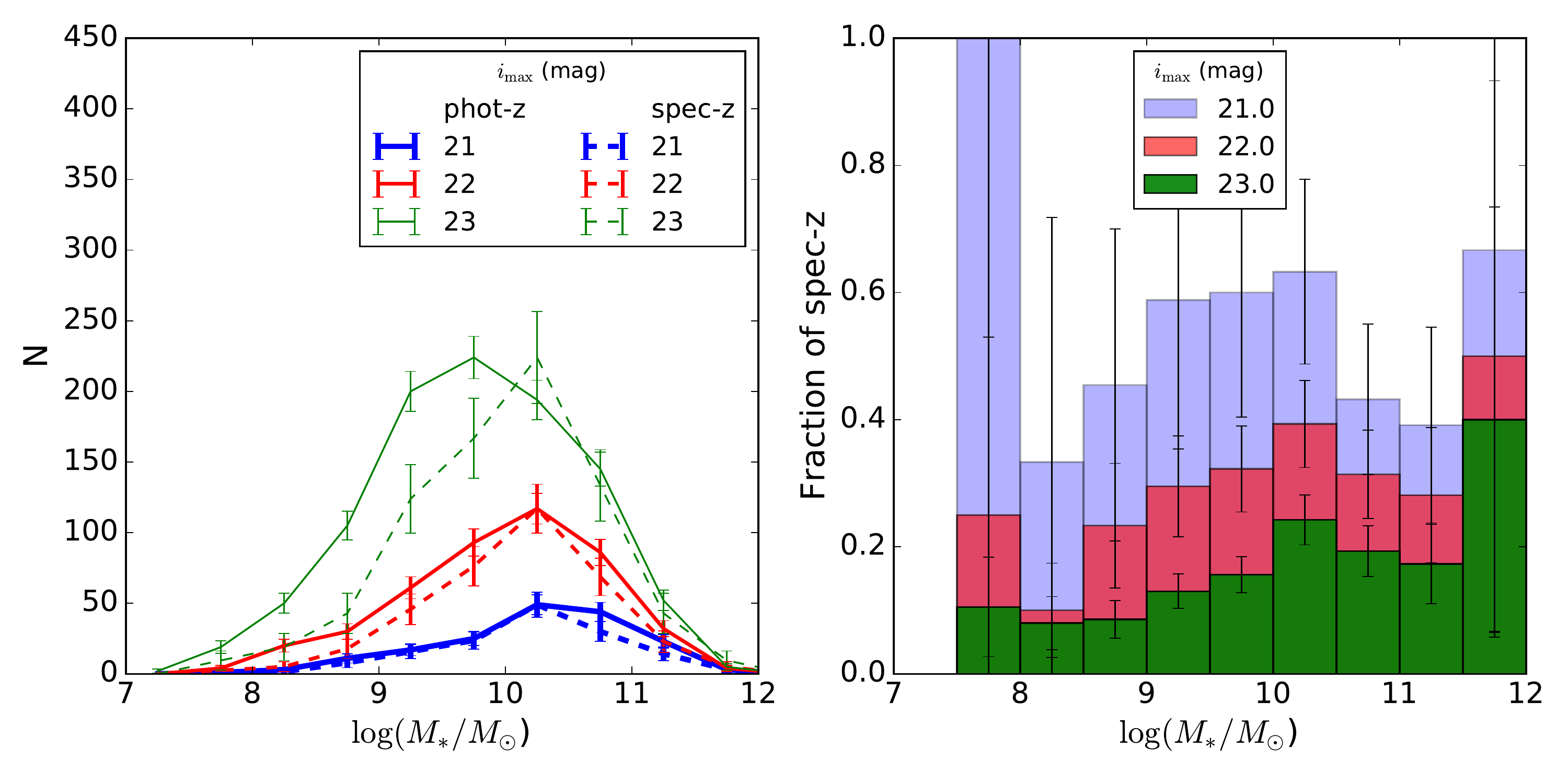}
	\caption{Characteristics of the spectroscopic sample for galaxies located at less than 6\arcmin\,from \HEofor.
		{\bf Left:} Number of galaxies as a function of the stellar mass for the photometric (solid) and spectroscopic (dashed) samples for three different cuts in magnitudes $i_{\rm max} = $ (21, 22, 23) mag (in resp. blue, red, green). To ease legibility, for each magnitude cut, the peak of the distribution of the spectroscopically confirmed galaxies has been normalized \cref{by a factor $n=(1.6, 2.5, 4.8)$} to match the corresponding peak \cref{(i.e. $i_{\rm max} = $ (21, 22, 23) mag)} of the photometric sample.
		 {\bf Right:} Fraction of spectroscopic redshifts as a function of the stellar mass for three different limiting magnitude $i_{\rm max} = $ (21, 22, 23) mag.} 
	\label{fig:Mass}
\end{figure}

We have also derived the fraction of objects with spec-z as a function of galaxy stellar mass. For that purpose, we use the mass and photometric redshifts (hereafter \emph{photo-z}) estimates obtained in \HOLI Paper III from multicolor optical photometry\footnote{Stellar masses derived using optical ($ugri$) + near infrared ($JHKs$) photometry were calculated only for the inner 2\arcmin\,around the lens due to the smaller field of view covered by the near infrared images. Consequently, we have only used stellar masses based on $ugri$ photometry. }. The left panel of Figure~\ref{fig:Mass} shows that our spectroscopic sample is not mass-biased down to $i \sim 22\,$mag. For a limiting magnitude $i \sim 23\,$mag, the photometric and spectroscopic distributions start to differ more significantly. This is because most of our spectroscopically confirmed galaxies have magnitudes $i < 22\,$ mag (Fig.~\ref{fig:magdistrib}). The right panel of Figure~\ref{fig:Mass} shows that the completeness of the spectroscopic sample is the highest (40-50\%) at the high mass end (i.e. $M_* \sim 10^{12}\, M_\odot$), even down to $i=23\,$mag, and remains above 30\% down to typically $M \sim 10^{10} M_\odot$.  

%-------------------------------------------------------------------------------

\section{Galaxy group identification}
\label{sec:groups}

Our main objective is the identification of groups located close in projection to \HEofor\, as they are the most likely to influence the time delay between the lensed images. This requires a method sensitive enough to allow the detection of low mass and compact groups but also of loose groups. Spectroscopy-based techniques are particularly well suited for this aim but demand adaptive selection criteria. In general, group candidates are first identified based on peaks in redshift space, and then group membership is refined based on the spatial proximity between candidate group galaxies. The latter is assessed either based on the aspect ratio between the velocity-space group elongation along the line of sight (the ``finger of God effect") and its transverse extension \citep[e.g.][]{Wilman2005, Munoz2013}, or on a proxy for the group virial radius \citep[e.g.][]{Calvi2011, Ammons2014}. \cref{After experimentation, we found that the use of the aspect ratio to assess group membership yields detection of a larger number of groups than using $R_{\rm vir}$, additional groups being often poor groups and possibly yet non virialized structures. We therefore used that selection criterion because it provides a more complete census of groups and allows us to conservatively estimate their impact on the gravitational lensing potential. For the sake of completeness, we report and discuss the results obtained with the virial radius criterion in Appendix~\ref{Appendix:methodII}. The general design of our algorithm is described below.}

\subsection{Group identification}
\label{subsec:algorithm}

Our strategy to identify galaxy groups consists of two main steps. The first step is building a trial group catalog. Following \cite{Ammons2014}, we identify group candidates simply based on peaks in redshift space. Potential group redshifts are detected by selecting peaks of at least 5 members in redshift space with redshifts grouped in bins of 2000 \kms (observed frame). Groups of less than 5 members are unlikely to play an important role in the lens analysis as they most likely have a low velocity dispersion \citep[i.e. $\sigma$ peaks at $\sim $100\,\kms, ][]{Robotham2011}. The operation is repeated after shifting the bin centers by half the bin-width to avoid missing a peak due to an inadequate binning. Then, the potential group members are selected as being galaxies located within $\pm $1500 \kms of the peak, corresponding roughly to three times the velocity dispersion of a group with $M_{\rm vir} \sim 10^{13.7}-10^{14}\,M_\odot$. Additional neighbor galaxies in redshift space are included in the group if they are located less than 1500 \kms from another candidate group member. A trial group catalog can then be constructed.  These conservative starting criteria are meant to enhance our sensitivity to small groups, which typically will have velocity dispersions of a few hundred \kms.  A bi-weight estimator \citep{Beers1990} is used to calculate the mean redshift and velocity dispersion of the group candidates. The groups centroid is determined as a luminosity weighted centroid \citep{Wilman2005, Robotham2011}.

The trial groups obtained are the starting point for the second step of our algorithm that measures the spatial separation between galaxies sharing similar redshifts, and iteratively refine the group properties. The procedure removes those galaxies that are too far in the outskirts of the group and/or in the tail of the group redshift distribution. The algorithm follows a methodology similar to that of \cite{Wilman2005} as described below. 
\begin{enumerate}
	\item The initial group redshift is derived from our trial group catalog. Because the selection criterion of the trial catalog is very conservative it largely overestimates the group velocity dispersion. In order to identify even small groups, we proceed like \cite{Wilman2005} and initially set $\sigma_{obs} = $\,500 \kms. That value is revised in subsequent iterations of the algorithm.  
	\item Galaxies that are more than $n$ times the group velocity dispersion from the group redshift are excluded. This corresponds to the following limit in redshift space: 
	\begin{equation}
	\delta z_{\rm max} = n \times \sigma_{\rm obs} / c, 
	\label{eq:dz}
	\end{equation}
	\noindent where $n=2$ is used, and $\sigma_{\rm obs}$ is the group velocity dispersion uncorrected for redshift measurement errors.  
%	Like \cite{Wilman2005}, we find that a two sigma clipping (i.e. $n=2$) yields 
	\item The maximum angular transverse extension of the group $\delta \theta_{\rm max}$ is derived assuming an aspect ratio $b = 3.5$ for the group, giving
	\begin{equation}
	\delta \theta_{\rm max} = 206265\arcsec \frac{c \times \delta z_{\rm max} } {(b\,(1 + z)\,H(z)\,D_{\theta}(z) ) }
	\label{eq:dtheta}
	\end{equation}
        where $D_{\theta}(z)$ is the angular diameter distance to redshift $z$.
	\item The angular separation between each galaxy and the \cref {$i$-band} luminosity weighted group centroid is derived and galaxies that have $\delta \theta < \delta \theta_{\rm max}$ and $|z - \bar{z}_{\rm group}| < \delta z_{\rm max}$ are kept as group members. If a galaxy lacks a reliable photometric measurement (this happens for about 5\% of the galaxies of our catalog), we do not use a luminosity weighting scheme for the galaxy centroid. \cref{This has no impact on the group detection but generally changes appreciably the group centroid. The difference in group centroid position has no significant impact on the cosmological analysis performed in \HOLI Paper V.}
	\item The observed group velocity dispersion $\sigma_{\rm obs}$ is recalculated using the gapper algorithm \citep{Beers1990} if the group contains fewer than 10 galaxies, and a bi-weighted estimator otherwise. This procedure is known to provide a less biased estimate of the velocity dispersion \citep{Beers1990, Munoz2013}. If during this iterative process the number of group members falls below 4, the standard deviation is used instead, as none of the other technique provides reliable estimate of $\sigma_{\rm obs}$ for a small number of objects. At the same time we also derive an improved group redshift using the bi-weight estimator, or the mean when we are left with fewer than 4 members.  
	\item A new centroid is redefined based on the new members, and a new group redshift $\bar{z}_{\rm group}$ is derived using a bi-weight estimator. The whole process (from ii) is repeated until a stable solution is reached. A solution is generally found after three to five iterations. 

\end{enumerate}

Once a stable solution is reached, the intrinsic velocity dispersion of the group (i.e. obtained after converting galaxy velocities to rest frame velocities using $v_{\rm rest} = c\,(z-\bar{z}_{\rm group})/(1+\bar{z}_{\rm group})$) is computed, removing in quadrature the average measurement error of the group galaxies from the (rest-frame) velocity dispersion \citep{Wilman2005}.

\subsection{Caveats}
\label{subsec:caveats}

\cref{The group detection depends to some extent on the choice of the parameters used in our iterative algorithm, in particular of the value of the aspect ratio $b$ and of the rejection threshold in velocity space (i.e. $n$ in Eq.~(\ref{eq:dz})). The fiducial values used for those parameters have been chosen based on those used in \cite{Wilman2005}. We experimented with different choices, including aspect ratio $b=11$ \citep[as found in some numerical simulations, e.g.][]{Eke2004}, rejection threshold $n=3$. We found that the fiducial value of $b$ tends to maximize the number of group members as well as the chance of detection of a group at a given peak in redshift space. The choice of rejection threshold at $n=3$ favors the identification of larger groups with multiple peaks in redshift space, suggesting that non-group members are included. }

%-------------------------------------------------------------------------------

\section{Environment and line of sight characteristics}
\label{sec:environment}

Individual galaxies located close in projection to the main lens, as well as more distant galaxy groups, can significantly modify the structure of the lensing potential. In such a case, they need to be included explicitly in the lens model \citep{McCully2016}. We show in Sect.~\ref{subsec:galaxies} the spectra of the five galaxies that yield the most important perturbations of the lens potential. In Sect.~\ref{subsec:groups} \& ~\ref{subsec:groupdiscuss}, we present and discuss the results of our search for important groups in the field of view of the lens. These results will be used in Sect.~\ref{sec:model} where we quantify the amplitude of the perturbation caused by these structures. 

\subsection{Nearby galaxies}
\label{subsec:galaxies}

\begin{figure*}
	% Figure generated with MXU_environment.makefigcompleteness2(cmulative=False, rmax=360)
	%\begin{tabular}{cc}
	%\includegraphics[width=0.3\linewidth]{he0435_field2.pdf} &
	\includegraphics[width=0.7\linewidth]{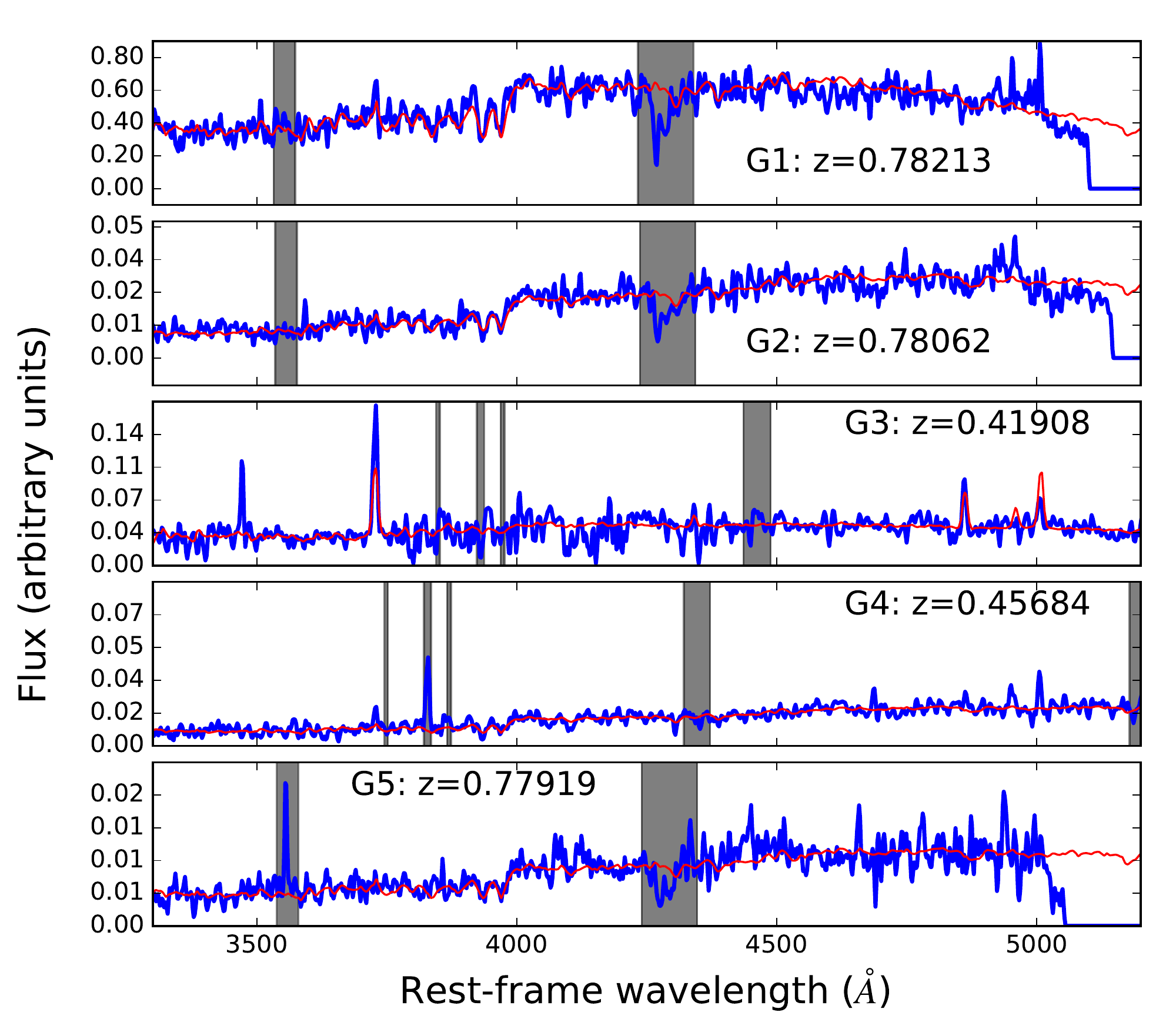}
	%\end{tabular}
	\caption{Rest-frame spectra of the galaxies G1-G5 (blue; see Fig.~\ref{fig:field} for identification) over-plotted with the best galaxy template (red) used to measure the redshift with the cross-correlation technique. For legibility, the spectra have been smoothed with a 5 pixels boxcar, and the templates have been multiplied by a third order polynomial to correct for uncertainties in the instrumental response. Gray bands indicate regions affected by sky subtraction problems.  }
	\label{fig:Spec}
	
\end{figure*}

\cref{When we initiated our spectroscopic follow-up, we were lacking color information for the galaxies in the field, precluding any selection based on photometric redshift or stellar mass. We therefore prioritized the follow-up based on the luminosity and projected distance to the lens (Sect.~\ref{subsec:target}). Five bright galaxies ($i < 22.5$\,mag) are detected at a projected distance of $r < \,$ 15\arcsec\,from the lens (Figure~\ref{fig:field}). Those galaxies, G1 to G5, were labeled G22, G24, G12, G21, G23 in \cite{Morgan2005}. Following the methodology proposed by \cite{McCully2016} and presented in Sect.~\ref{sec:model}, we have verified a posteriori (using photo-z and stellar mass, for galaxies without spec-z), that those galaxies are the most likely to influence substantially the modeling, the other faint galaxies detectable in the vicinity of the lens not yielding significant perturbation of the lens potential. Figure~\ref{fig:Spec} show the spectra and measured redshifts of galaxies G1 to G5}. Three of them are first time measurements. \cite{Chen2014} previously reported a redshift $z = 0.4188$ for G3 \citep[as well as ][]{Morgan2005}, and $z = 0.7818$ for G1. Those measurements are statistically compatible with ours. 
%\com{Those redshifts are NOT corrected for systematic error of z=0.0004 in FORS data.} 

The most important perturber (see Sect.~\ref{sec:model}), the galaxy G1, lies in the background of the lens at $z\sim 0.78$. It is potentially part of a small galaxy group of up to 4 spectroscopically identified members, including the two nearby galaxies G2 and G5. The galaxy G4 located $\sim$ 9\arcsec\, N-W of the lensing galaxy, is in the direct environment of the lens, and part of a larger group of galaxies at the lens redshift (see Sect.~\ref{subsec:groups}). The galaxy G3, at $z\sim$\,0.419, and located at $\delta \theta\,\sim$\,8\farcs6\,W-N-W from \HEofor, is the second most important source of perturbation of the gravitational potential after G1 (Sect.~\ref{sec:model}). The lens models presented in \HOLI Paper IV systematically include G1 using multi-lens plane formalism, while G2 to G5, which are found to impact less significantly the lens models due to their larger projected distance to the lens (see Sect.~\ref{sec:model}), are included in one of the systematic tests presented in that paper. 

% Figure generated with MXU_environment.figspec(polorder=3, plotrest=True)

\subsection{Groups: overview}
\label{subsec:groups}

\begin{figure}
	\includegraphics[width=1.0\linewidth]{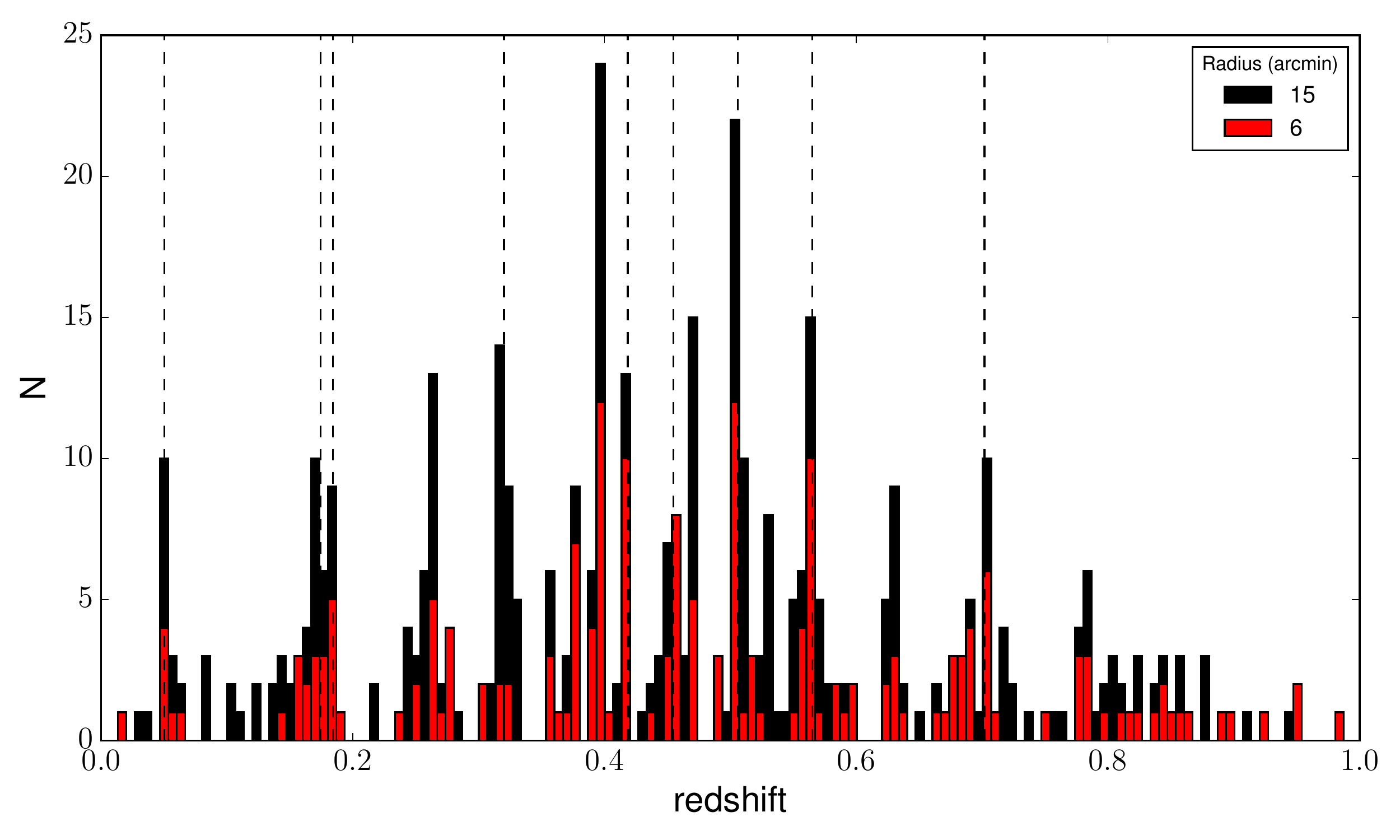}
	\caption{\cref{Redshift distribution of the objects within apertures of 6\arcmin (red) and 15\arcmin (black) centered on the lens. The redshifts of the groups identified with our algorithm (Tab.~\ref{tab:groups}) are shown as vertical dashed lines. Note that the height of the peaks changes when the bin center is offset by half the bin width. This ``redshift-phase'' effect is accounted for in our group detection algorithm (Sect.~\ref{subsec:algorithm}).}}
	\label{fig:zdistrib}
\end{figure}

Important perturbations to the lens potential are not only caused by individual galaxies, but can also be produced by more distant and massive groups along the line of sight, or by a group at the lens redshift. 
In order to flag those potential perturbers, it is mandatory to be able to detect even low-mass groups along line of sight that have their centroid located in projection within a few arcminutes from the lens, namely the virial radius of a typical group at the lens redshift. Owing to the spectroscopic completeness of our sample, we first apply our group finding algorithm (Sect.~\ref{subsec:algorithm}) to a region of 6\arcmin\, radius around the lens, where we have spectroscopically identified 1/3 of the galaxies down to $i=22$\, mag. \cref {Out of the ten peaks in redshift space observed in that range (Fig.~\ref{fig:zdistrib})}, seven lead to a group identification with our iterative procedure (Table~\ref{tab:groups}). 

By limiting the group search to a small field, we may underestimate the group richness, and in particular miss an important fraction of the galaxies lying in rich groups with a projected center significantly offset with respect to the lens position on the sky. It is therefore necessary to expand our search up to the largest radius available, namely 15\arcmin, in order to identify those structures. At those radii, the spectroscopic completeness drops significantly, but this is compensated by our quest for only the richest groups\footnote{Note that small groups at low redshifts can potentially have their centroid close in projection to the lens while being detectable only based on large area search due to their higher angular virial radius. \cref{However, the redshift difference between those groups and the main lens ensures a small effective impact on the lens potential, see Sect.~\ref{sec:model}}}.
In addition, because the group properties are particularly uncertain when the number of galaxy members is small and spectroscopic incompleteness high, \cref{we search for groups within 15\arcmin\, of the lens only around peaks in redshift space of at least 10 galaxies}. This choice is guided by the results obtained at smaller radii where group properties are more robustly retrieved above 10 galaxies. It is also above this threshold that our estimator of the group velocity dispersion is expected to be the most accurate \citep{Beers1990}. \cref{From the ten peaks found in redshift space (Fig.~\ref{fig:zdistrib}), only six are found to be associated with groups of at least five members (Table~\ref{tab:groups}). Two of these groups were undetected when we limited our search to a maximum separation of 6\arcmin\, from the lens.}

A complementary approach would be to search for groups based on photometric redshifts. Although, this technique should allow the detection of overdensities of galaxies with reasonable efficiency \citep{Williams2006, Gillis2011}, it would not allow us to characterize the group properties with sufficient accuracy due to the too large uncertainty on individual photometric redshifts ($\sigma_z = 0.07$), and of a small bias at the level $\sigma^{\rm sys}_z = 0.007$.

In total, we have identified 9 groups. Their properties are listed in Table~\ref{tab:groups}, their spatial and redshift distribution are shown in Fig.~\ref{fig:maingroups}, and an estimate of their virial mass and radius is provided in Appendix~\ref{appendix:Mvir}. \cref{The redshift distribution and spatial extension of two groups, at $z=0.5059$ and $z=0.5650$, suggest that these groups could be bimodal (Fig.~\ref{fig:maingroups}), namely constituted of two or more subgroups not identified as seperated structures by our algorithm. The use of the virial radius to identify groups (Appendix~\ref{Appendix:methodII}) yields group detection at the same redshifts but for two groups ($z=0.4185$ and $z=0.7019$). The group properties are compatible between the two selection criteria for all commonly identified groups except the possibly bimodal groups, and the group at $z\sim 0.32$. These differences are discussed in Appendix~\ref{Appendix:methodII}. We also note that {\emph {not}} using luminosity weighting centroid yields detection of two more group candidates: a group at z=0.3976 ($\sigma_{\rm int}=143\pm51$\,\kms, FOV=6\arcmin), and a group at $z=0.5651$ ($\sigma_{\rm int}=259\pm75$\,\kms, $N=5$, FOV=15\arcmin). }
	
Error bars on the velocity dispersion and centroid have been derived using a bootstrapping approach. This consists in constructing 1000 samples of each group, each sample having the same richness as the fiducial group, but with members randomly chosen among the fiducial ones (repetitions being allowed). When constructing the samples, we have independently bootstrapped the positions, redshifts and luminosity of the galaxies, and derived the group properties in the same way as for the real group (but we did not apply our iterative algorithm on the sample). The final uncertainty on the scrutinized group property is the standard deviation of the bootstrap distribution.  

% Figure generated directly in the notebook http://localhost:8888/notebooks/work/HOLICOW/HE0435_MXU/MXU_environment.ipynb

\begin{table*}
	% Based on tab_mg_360_m1.tex and tab_mg_360_m3.tex generated by MXU_environment.maingroups(rmax=360) 
	% Groups below do include a statistical correction of redshift
	\caption{Properties of the groups identified in the field of view of \HEofor. The columns are the group redshift, the number of spectroscopically identified galaxies in the group, the group intrinsic velocity dispersion (rounded to 10\,\kms\, maximum precision) and 1$\sigma$ standard deviation from bootstrap, the group centroid, bootstrap error on the centroid, projected distance of the centroid to the lens, median flexion shift $\log(\Delta_3 x (arcsec))$ and $\sigma$ standard deviation from bootstrapping (Sect.~\ref{sec:model}). \cref{The last column indicate for which field of view the group is detected. The properties we display correspond to the field of view marked in bold.}}
	\label{tab:groups}
	\centering
	\begin{minipage}{\linewidth}
		\centering
	\begin{tabular*}{0.8\linewidth}{lrcccccc}
		\hline
		$\bar{z}_{\rm group}$ & N & $\sigma_{int}$ $\pm$ err& $\alpha_{\rm ctr}$, $\delta_{\rm ctr}$ & err($\alpha_{\rm ctr}$, $\delta_{\rm ctr}$) &  $\Delta\theta$ & $\log(\Delta_3 x) \pm $ err & FOV  \\
		& & \kms & deg &  arcmin &  arcsec   & $\log(arcsec)$ & arcmin \\
\hline 
0.0503& 9& 163 $\pm$30 & 69.619870, -12.349930 & 1.69, 1.46& 303.6& -6.98 $\pm$ 0.75 &  {\bf {15}} \\ %sg0 m1 , 900 
0.1744& 6& 450 $\pm$ 100 & 69.548372, -12.280593 & 1.69, 2.00& 53.8& -4.99$\pm$ 1.41 & {\bf {6}} \\ %mg0, m1, 360    
0.1841& 5& 400 $\pm$ 100 & 69.620032, -12.310350 & 1.23, 1.11& 220.3& -6.06$\pm$ 1.35 &  {\bf {6}}  \\ % 360 mg1 m1  V
0.3202& 17& 470$\pm$ 70 & 69.535728, -12.363713 & 2.66, 0.96& 289.9& -5.96$\pm$ 0.45 &  {\bf {15}} \\  % sg4 , 900, m1 
0.4185$^{\dagger}$& 10& 280$\pm$ 70 & 69.549725, -12.301072 & 1.09, 0.90 & 65.5 & -5.58$\pm$ 0.87 & {\bf {6}}, 15 \\  % mg4, m1, 360 arcsec  V
0.4547& 12& 470$\pm$ 100 & 69.550841, -12.272258 & 0.65, 0.64& 67.1& -4.11$\pm$ 1.07 &  6, {\bf {15}} \\  % mg5 , 900arcsec, m1  V
0.5059$\ddagger$& 20& 450$\pm$ 60 & 69.607588, -12.242494 & 1.12, 0.65& 227.7& -6.01$\pm$ 0.33 &  6, {\bf {15}}  \\ % mg6, 900; m1 same with virial but 9 members; different centroid than 6arcmin -> bimodal
0.5650$^\ddagger$& 9& 330$\pm$ 60 & 69.571243, -12.281514 & 0.31, 1.22& 38.8& -5.29$\pm$ 0.91 & {\bf {6}}, 15 \\ %mg7, 360, m1 ;    different w. virial and different from 6arcmin -> bimodal
0.7019& 5& 170$\pm$ 60 & 69.555481, -12.282284 & 0.91, 0.57& 29.3& -6.81$\pm$ 1.38 &  {\bf {6}} \\  % mg8, m1, 360  Only aspect-ratio	
\hline
\end{tabular*}
\begin{flushleft}
{\small 
	{\bf Note:} ${\dagger}$ Galaxy members drop to 8 if a radius of 15\arcmin\, is considered. The centroid location does not change but the velocity dispersion drops to $\sigma=233\pm63 $\kms. $\ddagger$ Possibly bimodal groups constituted of two (or more) sub-groups. \\
}
\end{flushleft}

\end{minipage}

\end{table*}

\begin{figure*}
	% figures Generated with MXU_environment.maingroups(rmax=360, or rmax = 900 ; subgroup=True for sg)
	
	\begin{tabular}{cc}
		\setcounter{subfigure}{0}
     \subfloat[a][z=0.0503]{\includegraphics[ scale=0.5]{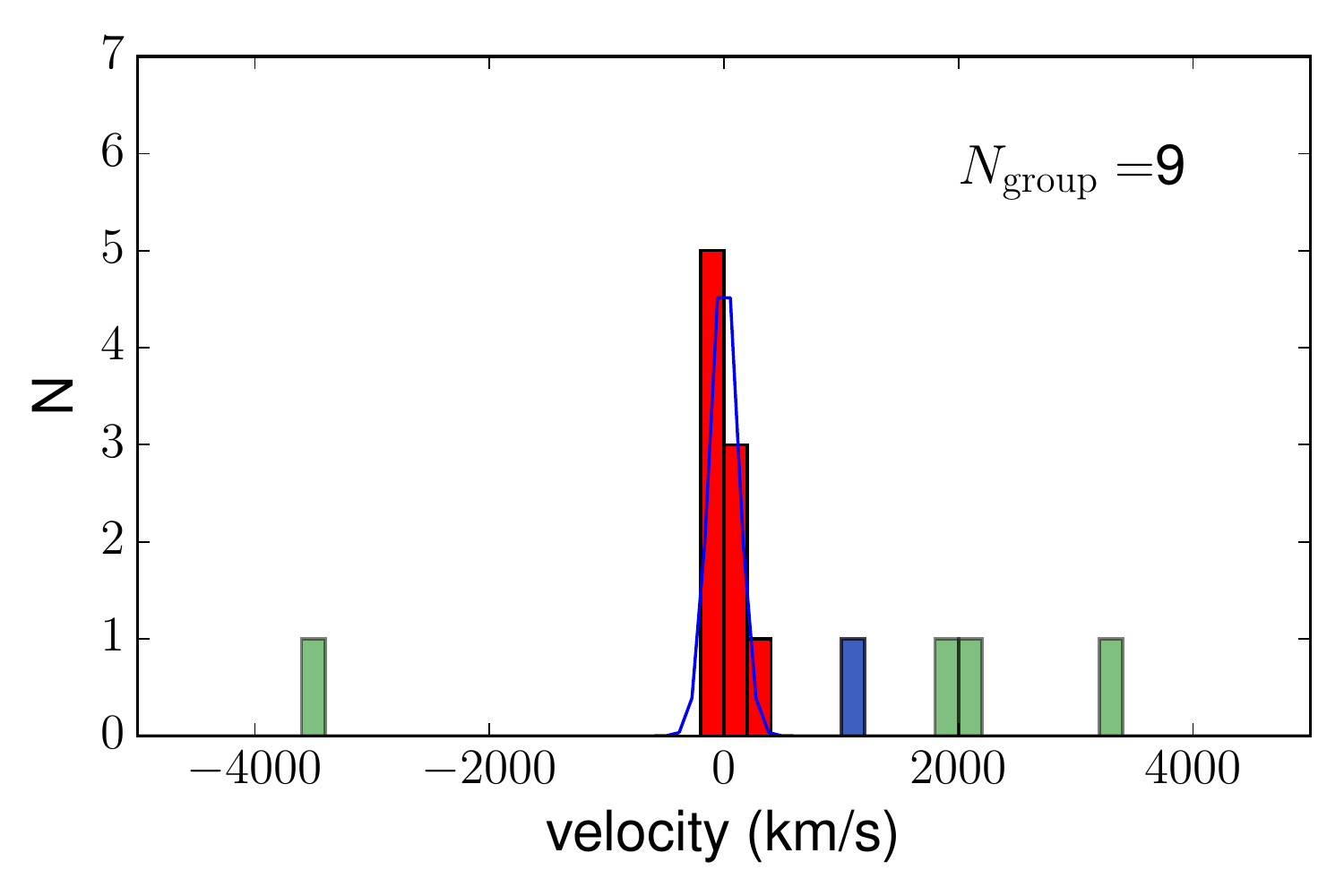}}
&	 \subfloat{\includegraphics[scale=0.4]{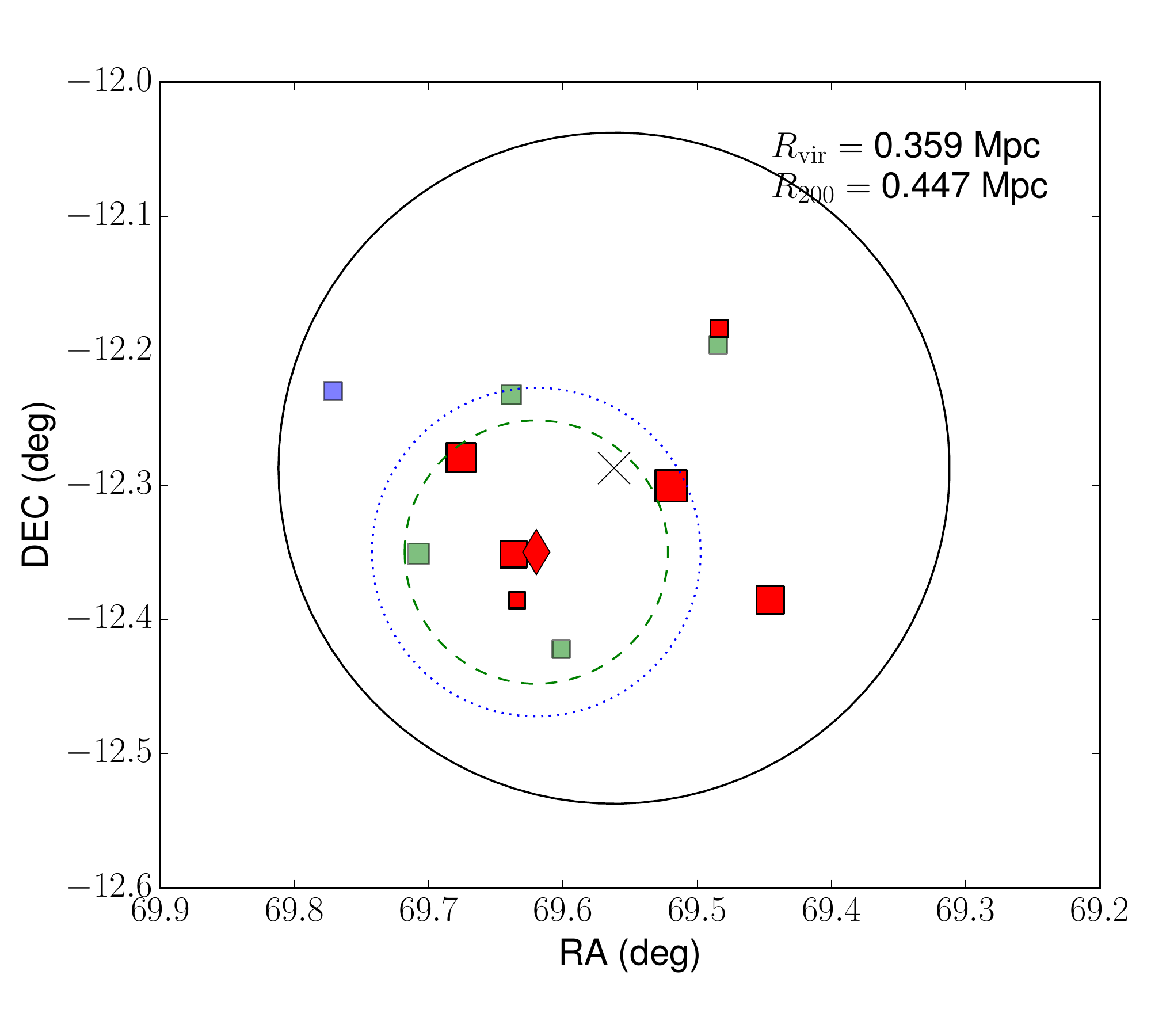}} \\ 	
%	&	\subfloat{\includegraphics[scale=0.4]{HE0435_prop_sg0_900_m1.pdf}}  \\
			\setcounter{subfigure}{1}
\subfloat[b][z=0.1744]{\includegraphics[scale=0.5]{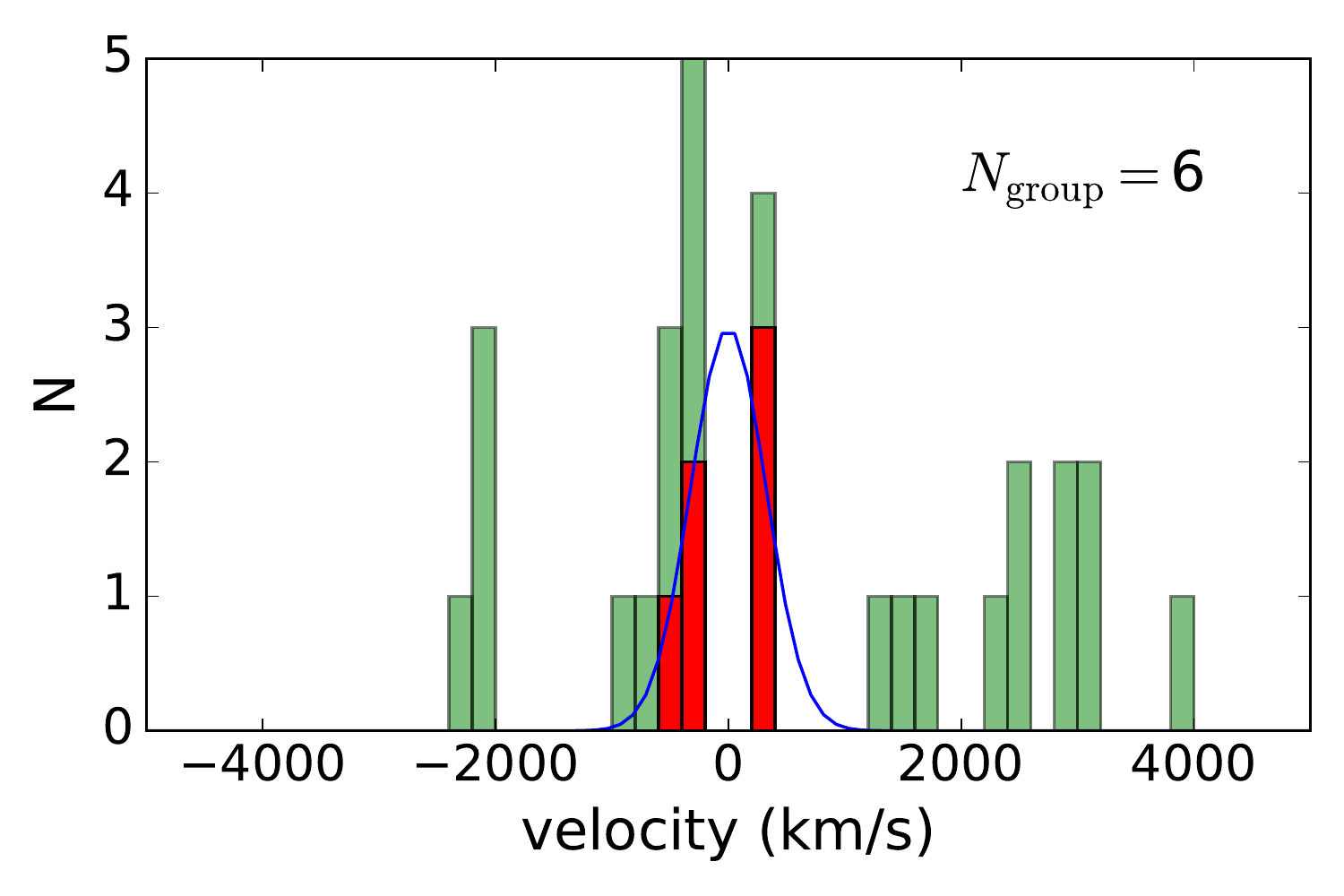}}
&	 \subfloat{\includegraphics[scale=0.4]{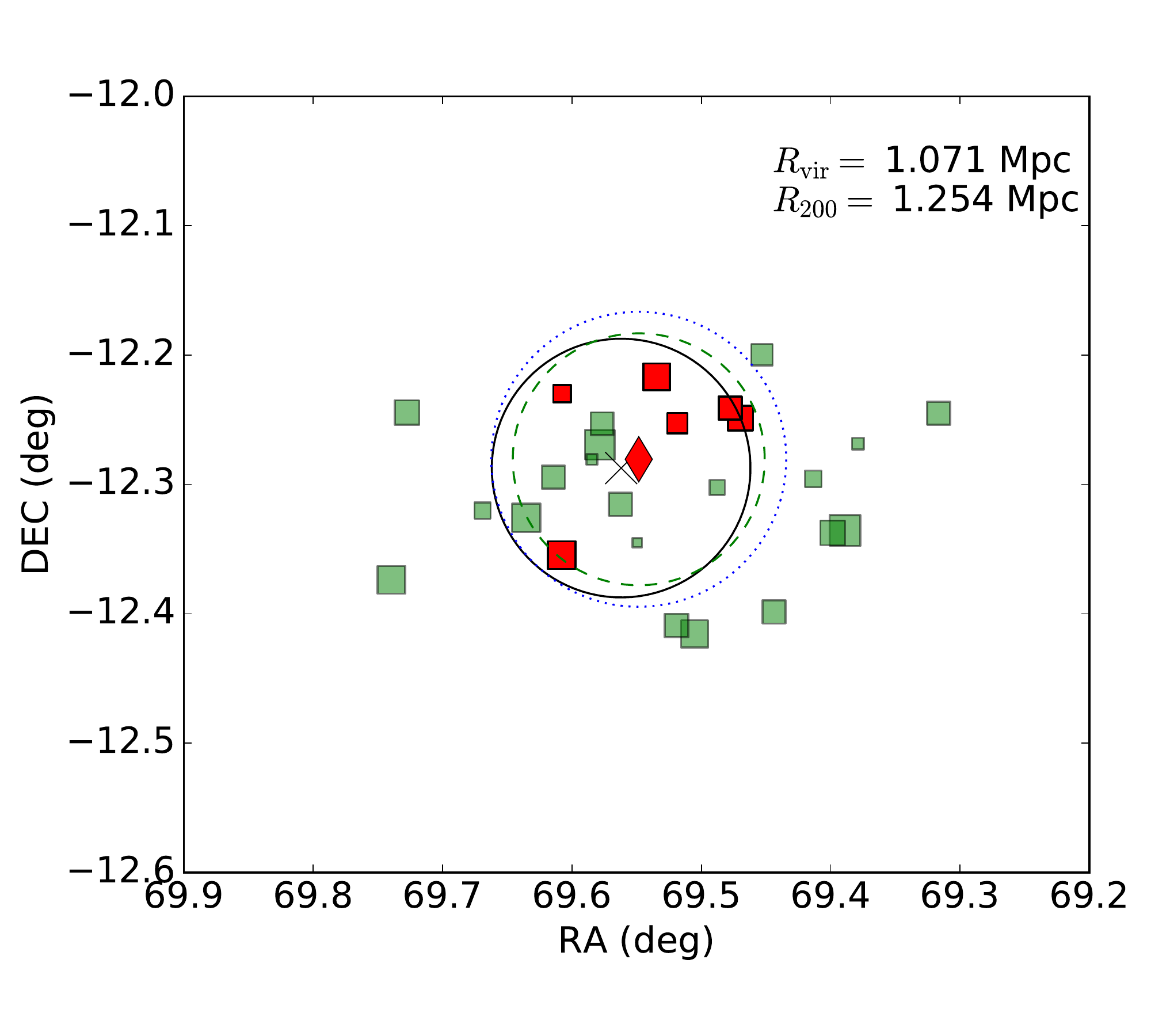}} 
%		&	\subfloat{\includegraphics[scale=0.4]{HE0435_prop_mg0_900_m3.pdf}}  \\

	\end{tabular}

	\caption{{Groups identified in the field of \HEofor:} For each redshift, the distribution of (rest-frame) velocities of the group galaxies identified spectroscopically is shown (left panel) together with a Gaussian of width equal to the intrinsic velocity dispersion of the group. Bins filled in red correspond to galaxies identified as group members, in blue as interlopers in redshift space, and in green as non-group members. The right panel shows the spatial distribution of the galaxies with a redshift consistent with the group redshift. The positions of the lens, group centroid and galaxies at $\sim \bar{z}_{\rm group}$ are indicated with a cross, diamond and square respectively. The size of the symbol is proportional to the brightness of the galaxy, and color code is the same as for the left panel. The solid black circle and blue-dotted (green-dashed) circles show the field used to identify the group, and a field of radius $r\sim 1 \times R_{\rm vir}$ ($r\sim 1\times R_{200}$). }
	\label{fig:maingroups}
\end{figure*}

\begin{figure*}
	\begin{tabular}{cc}
		\setcounter{subfigure}{2}
        \subfloat[c][z=0.1841]
      {\includegraphics[scale=0.5]{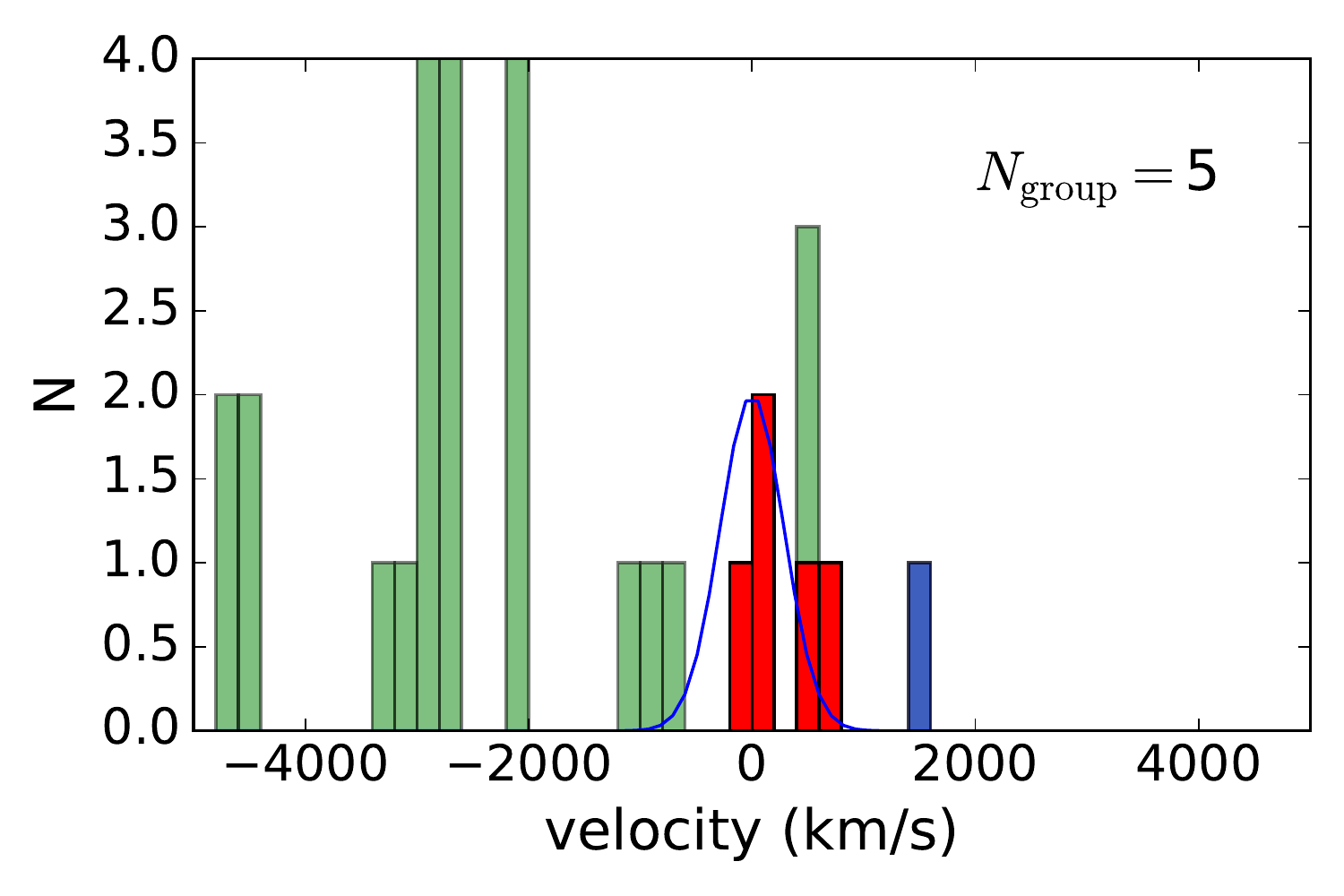}} 
		& \subfloat{\includegraphics[scale=0.4]{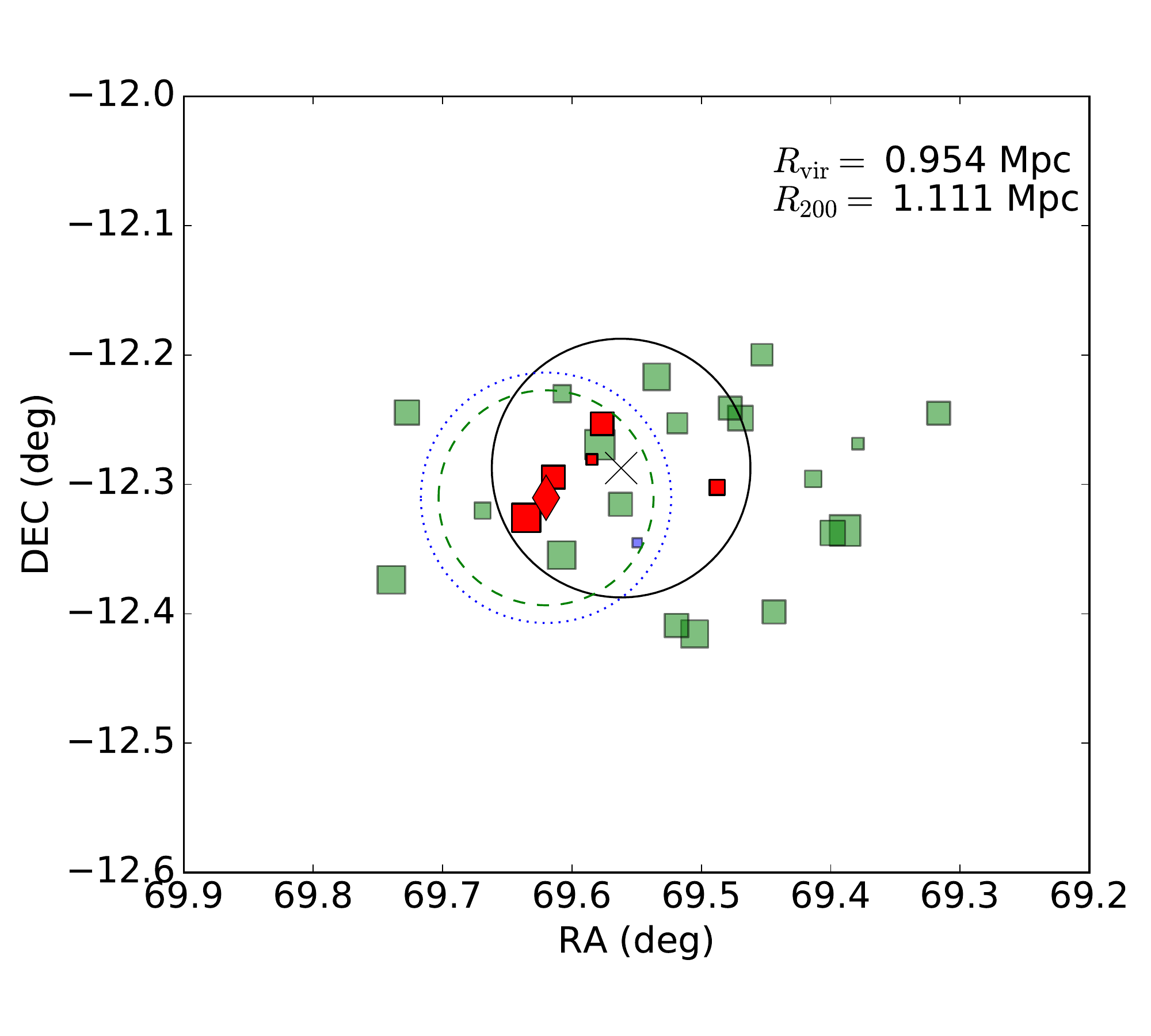}} \\ 	
%		&	\subfloat{\includegraphics[scale=0.4]{HE0435_prop_mg1_360_m1.pdf}}  \\
		\setcounter{subfigure}{3}
	   \subfloat[d][z=0.3202]{\includegraphics[scale=0.5]{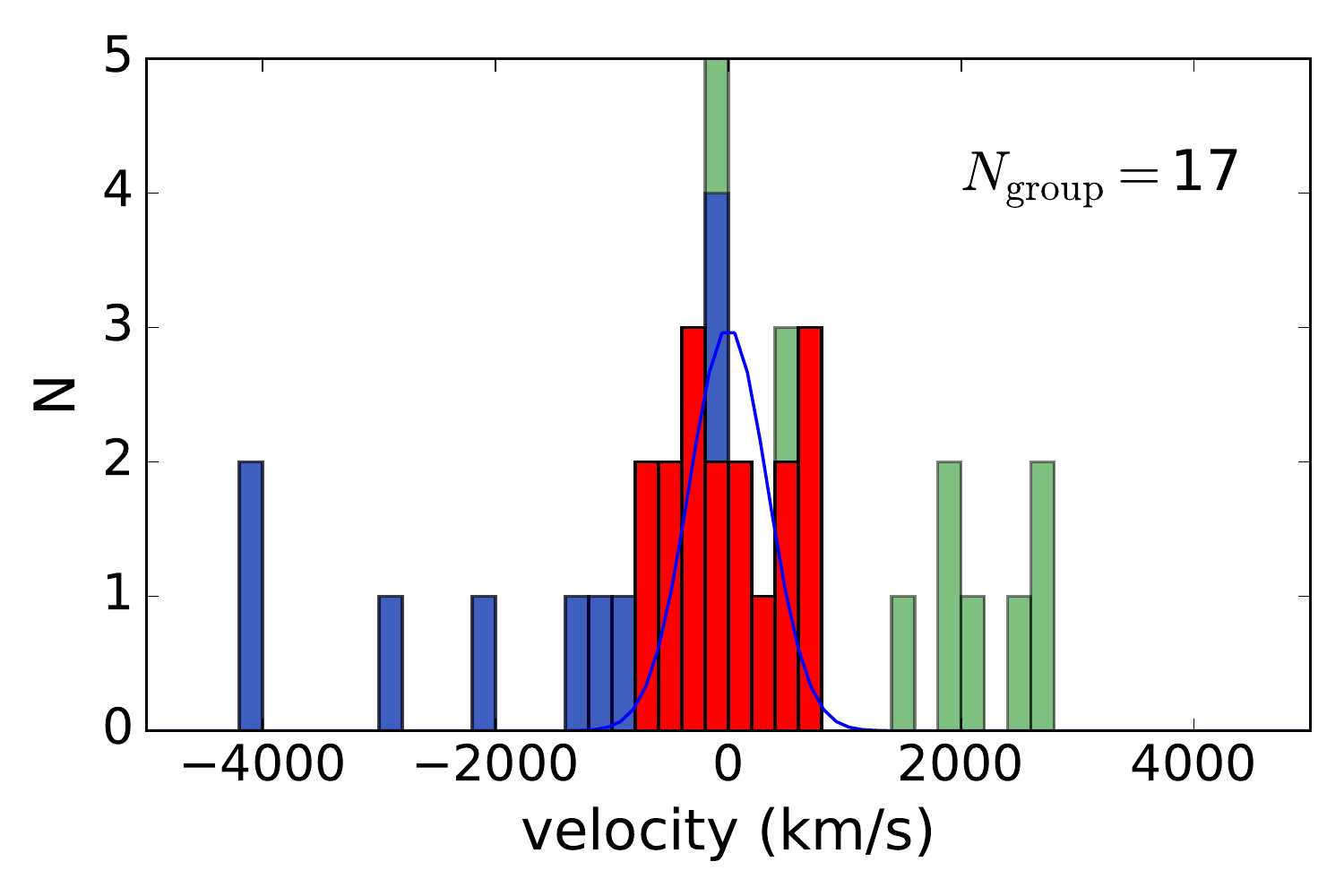}}
	   & \subfloat{\includegraphics[scale=0.4]{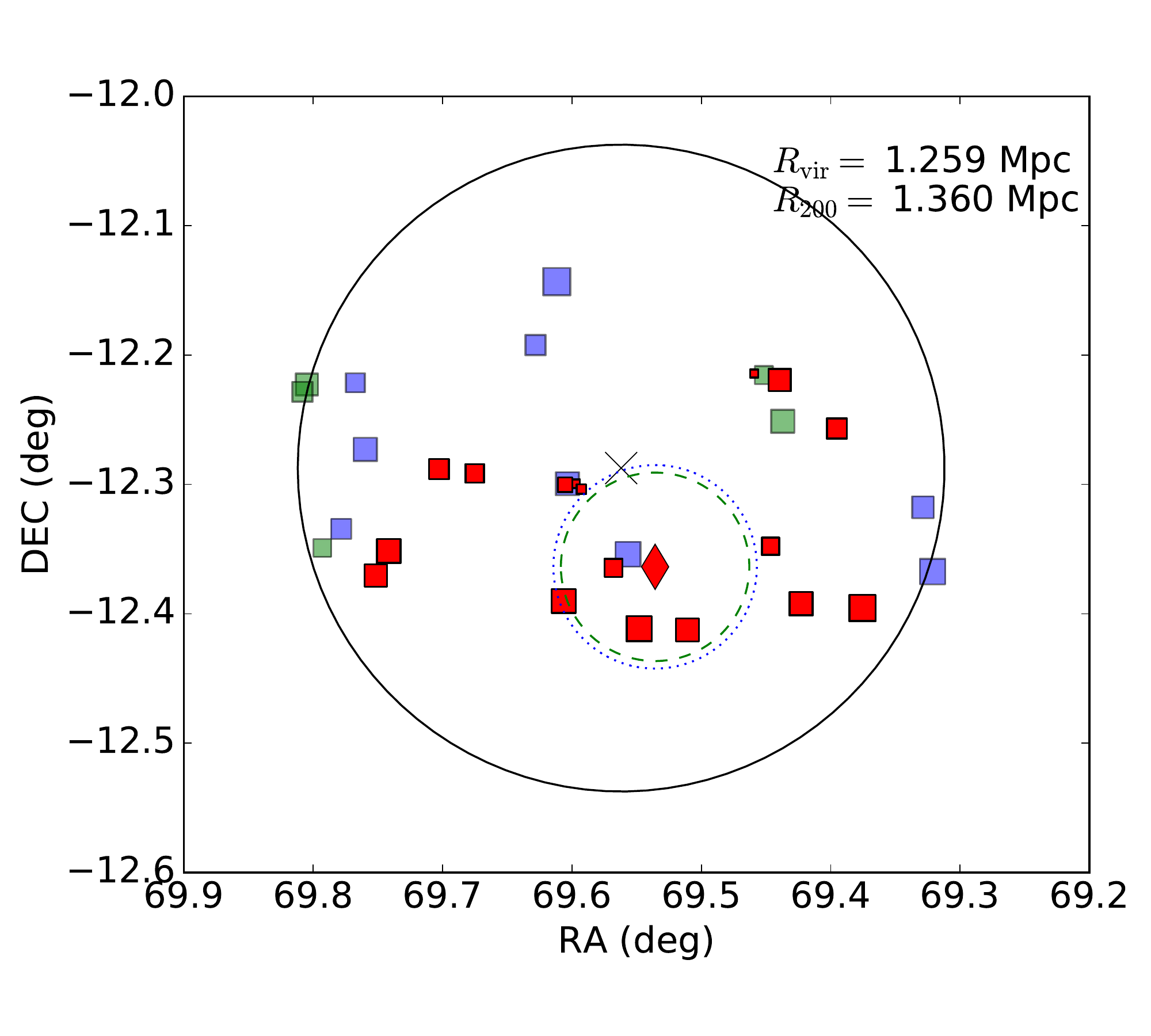}} \\ 	
	   %	&	\subfloat{\includegraphics[scale=0.4]{HE0435_prop_sg4_900_m1.pdf}}  \\
	\end{tabular}
	\setcounter{figure}{6}
	\caption{continued. }
\end{figure*}
		
\begin{figure*}
	\begin{tabular}{cc}
	\setcounter{subfigure}{4}
 \subfloat[e][z=0.4185]{\includegraphics[scale=0.5]{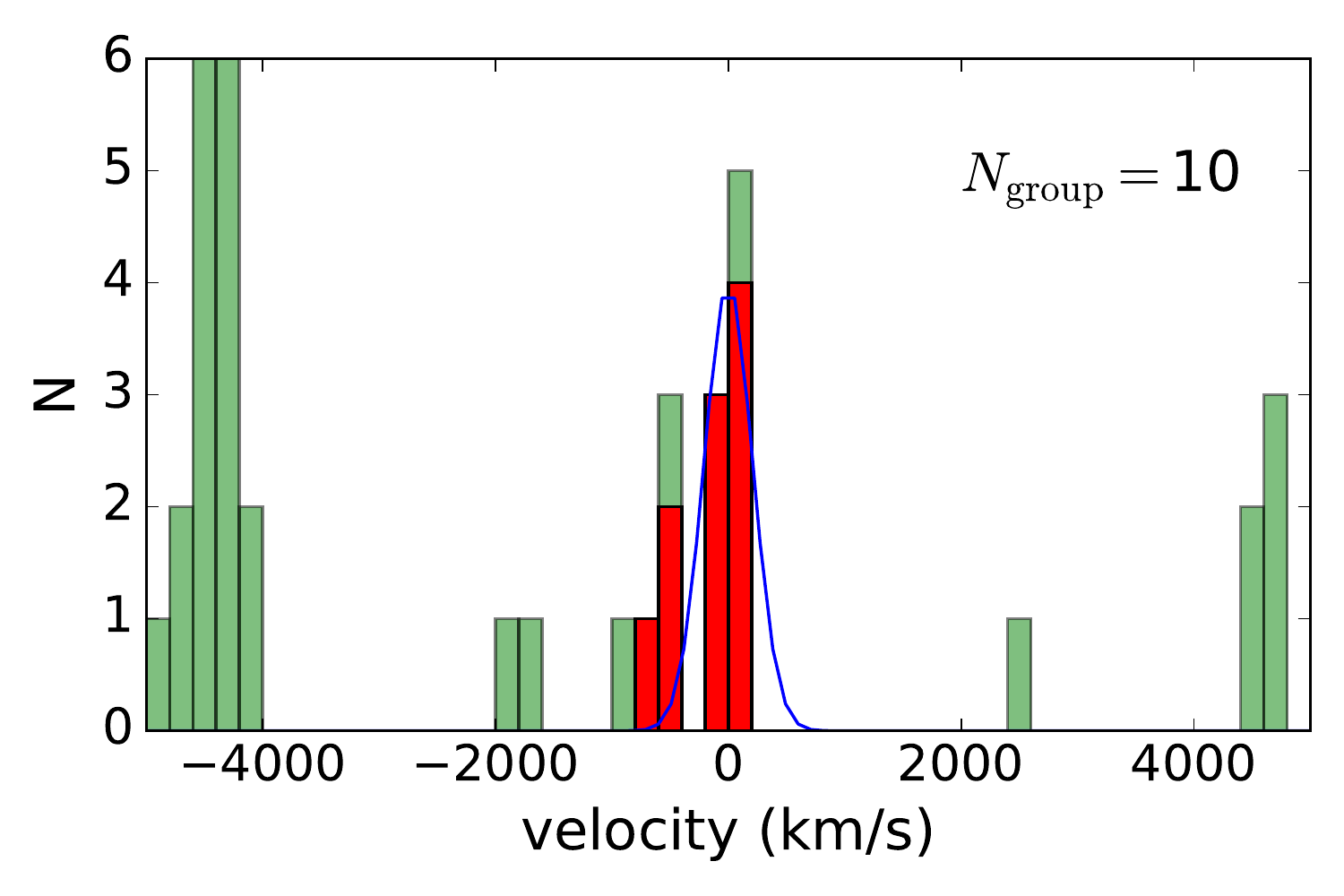}}
 & \subfloat{\includegraphics[scale=0.4]{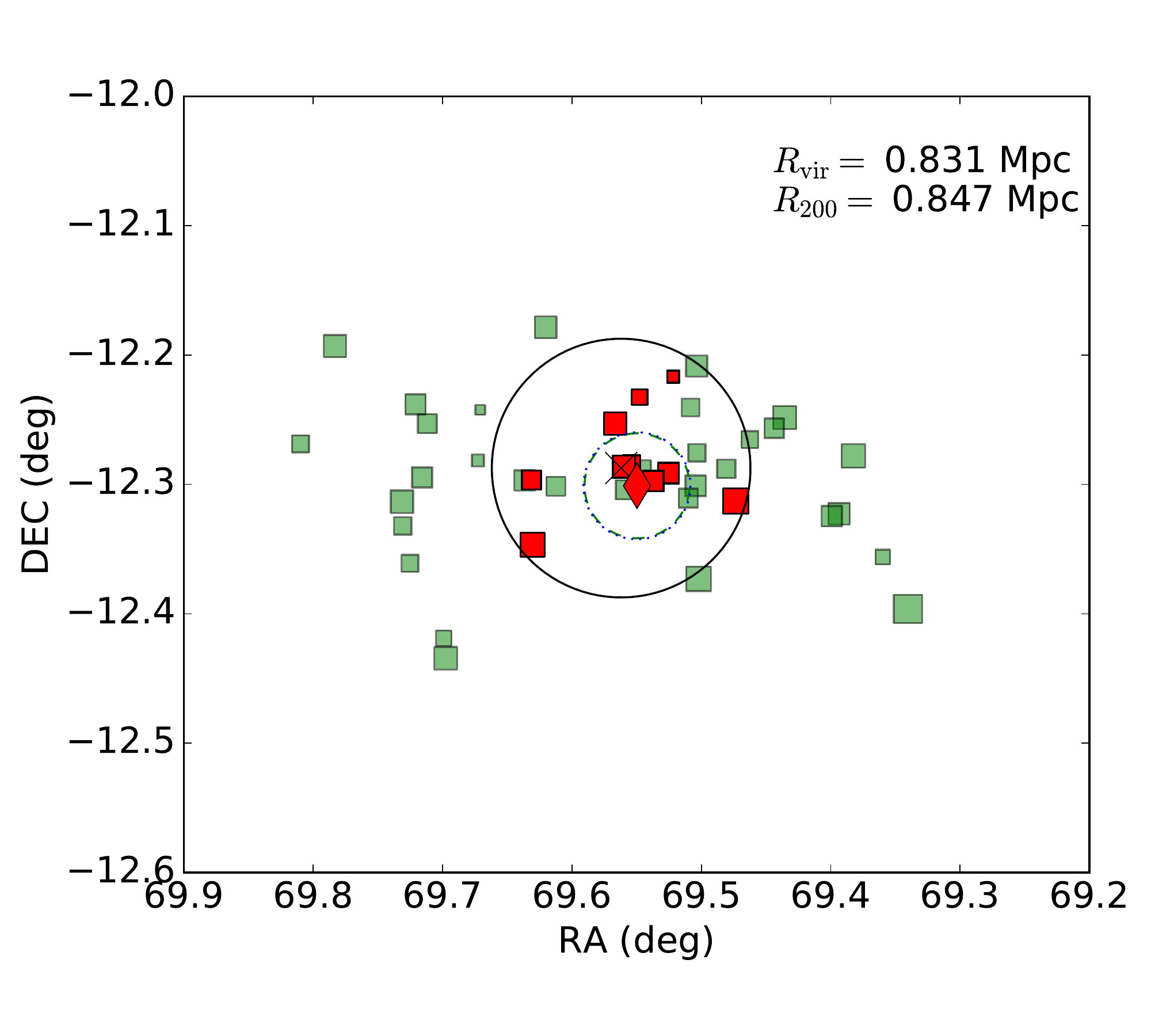}} \\ 	
 %		&	\subfloat{\includegraphics[scale=0.4]{HE0435_prop_mg4_360_m1.pdf}} 
\setcounter{subfigure}{5}
      \subfloat[f][z=0.4547]{\includegraphics[scale=0.5]{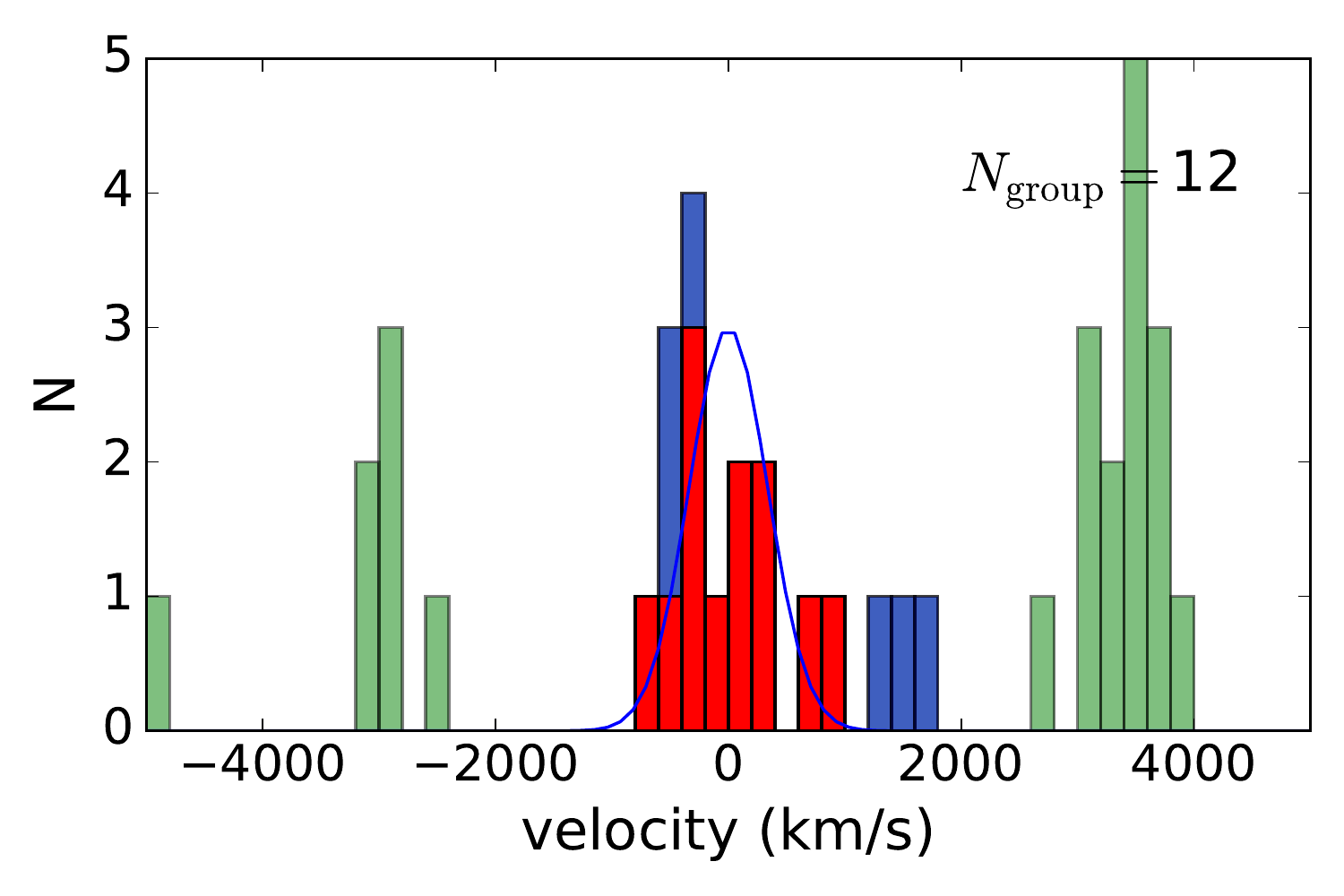}}
      & \subfloat{\includegraphics[scale=0.4]{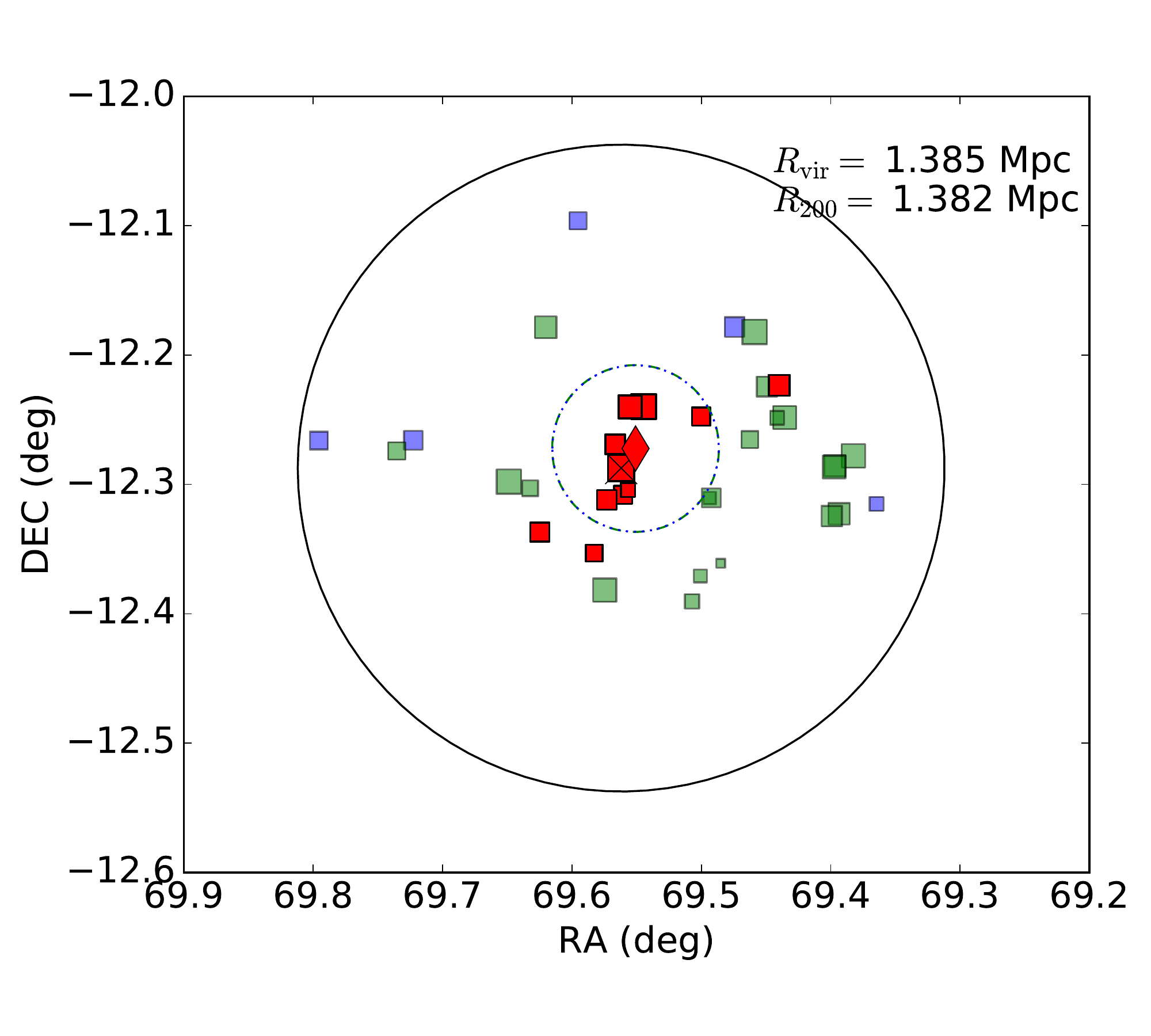}}  \\
      %		&	\subfloat{\includegraphics[scale=0.4]{HE0435_prop_mg5_900_m1.pdf}}  \\
\end{tabular}

\setcounter{figure}{6}
\caption{continued.}
\end{figure*}

\begin{figure*}
	\begin{tabular}{cc}
		\setcounter{subfigure}{6}
	\subfloat[g][z=0.5059]{\includegraphics[scale=0.5]{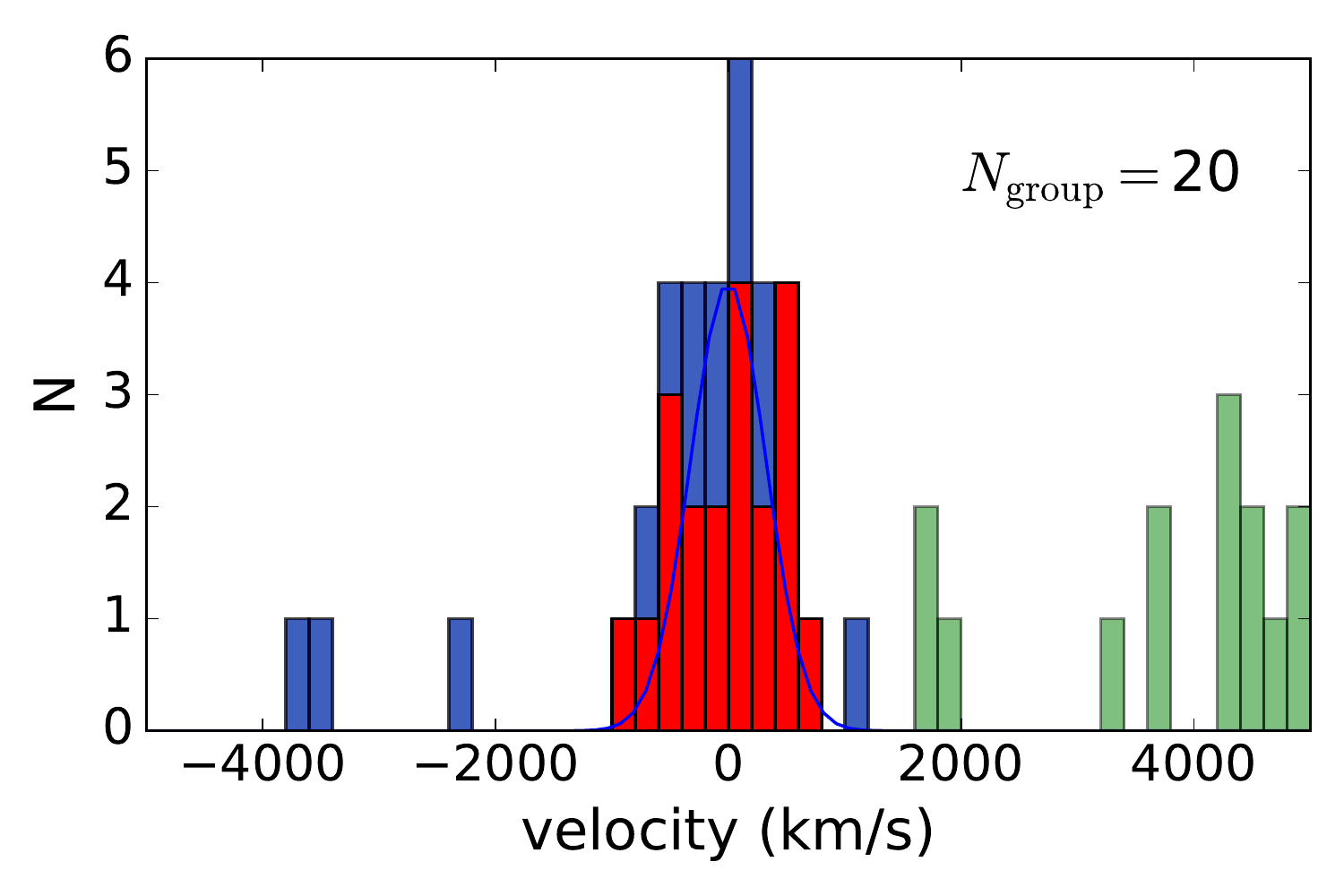}}
	& \subfloat{\includegraphics[scale=0.4]{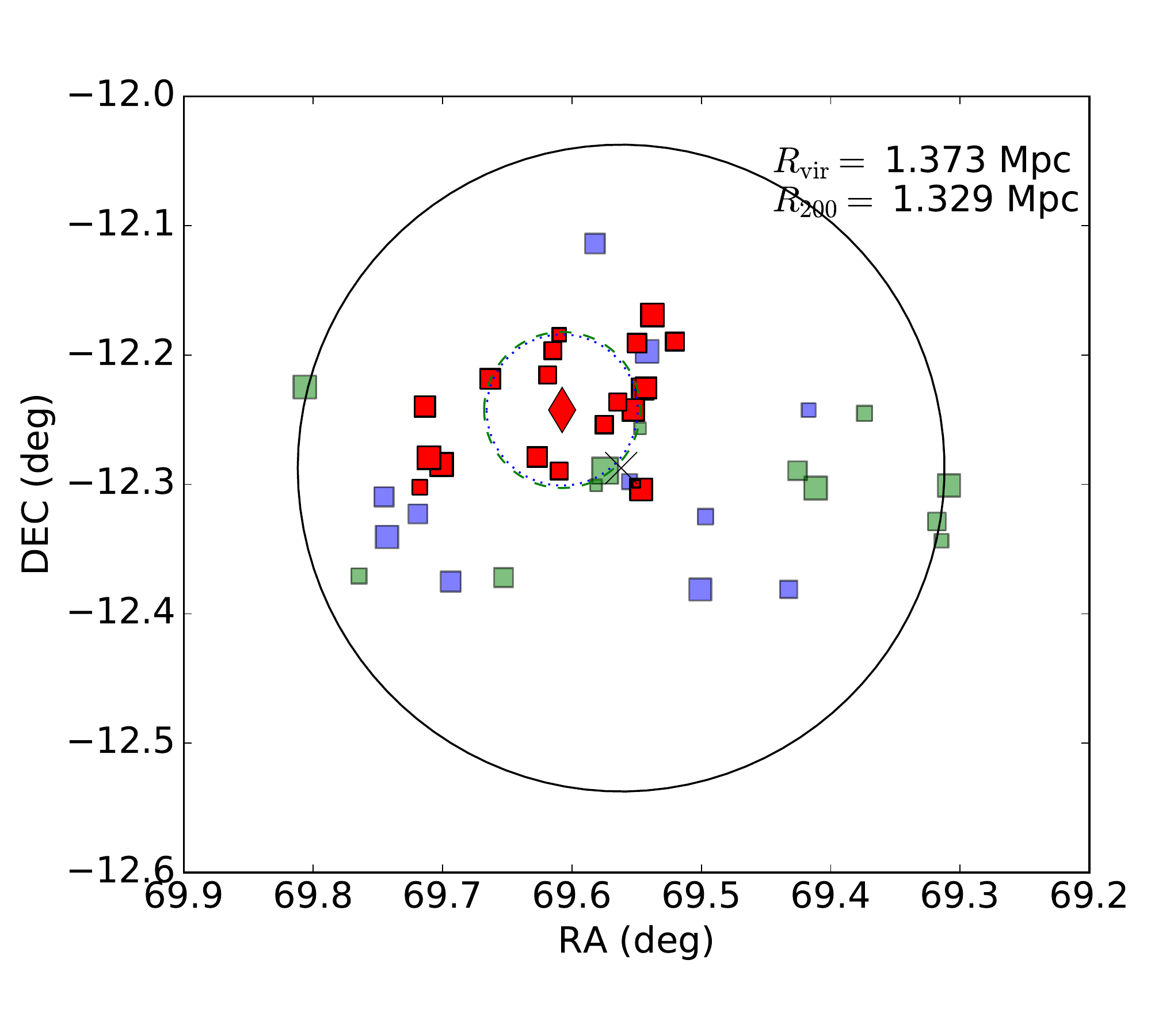}} \\
	%	&	\subfloat{\includegraphics[scale=0.4]{HE0435_prop_mg6_900_m1.pdf}}  
\setcounter{subfigure}{7}
     \subfloat[h][z=0.5645]{\includegraphics[scale=0.5]{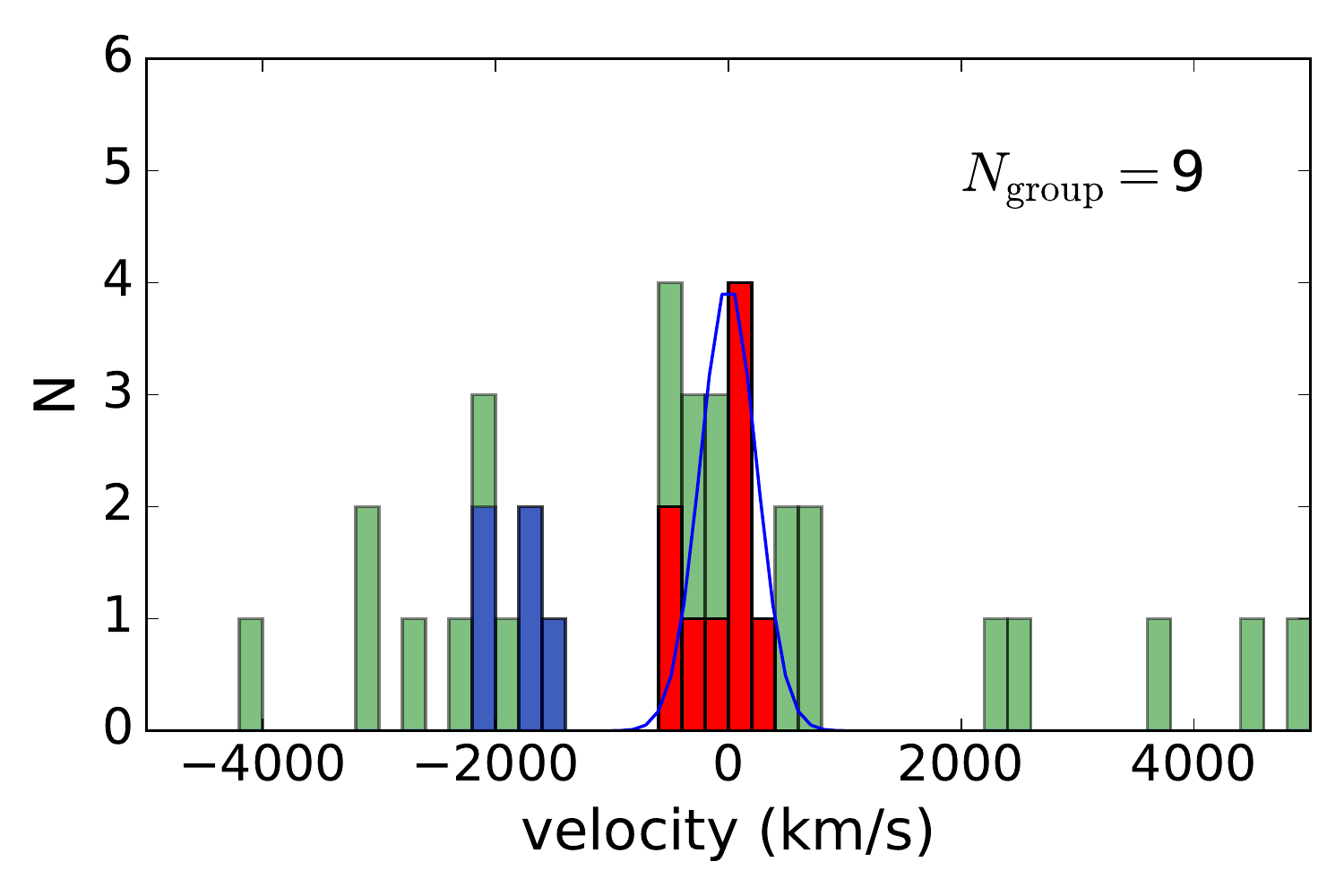}} 
     & \subfloat{\includegraphics[scale=0.4]{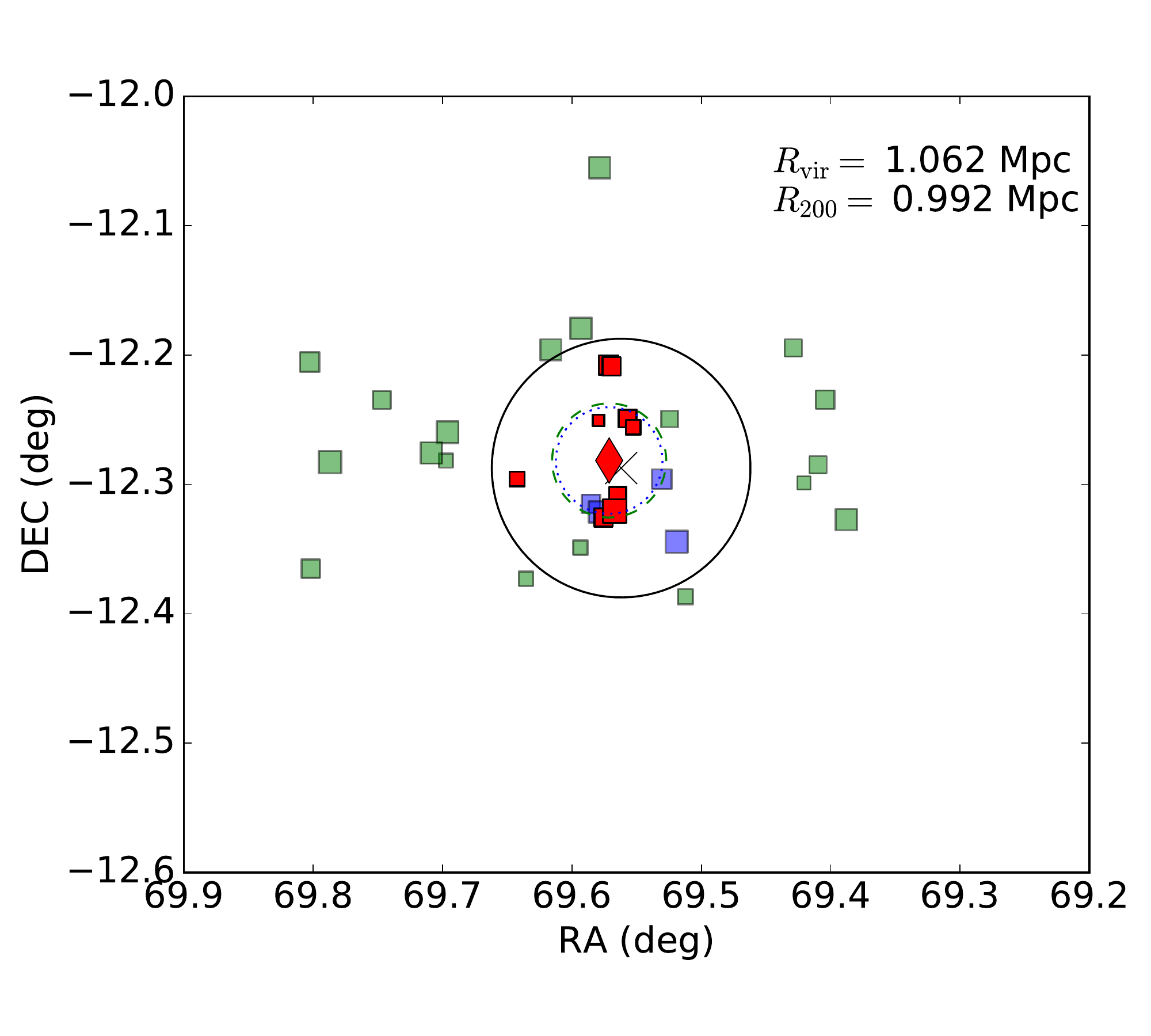}} \\
     %	&	\subfloat{\includegraphics[scale=0.4]{HE0435_prop_mg7_360_m1.pdf}}  \\

\end{tabular}

\setcounter{figure}{6}
\caption{continued.}
\end{figure*}

\begin{figure*}
	\begin{tabular}{cc}

	\setcounter{subfigure}{8}
     \subfloat[i][z=0.7019]{\includegraphics[scale=0.5]{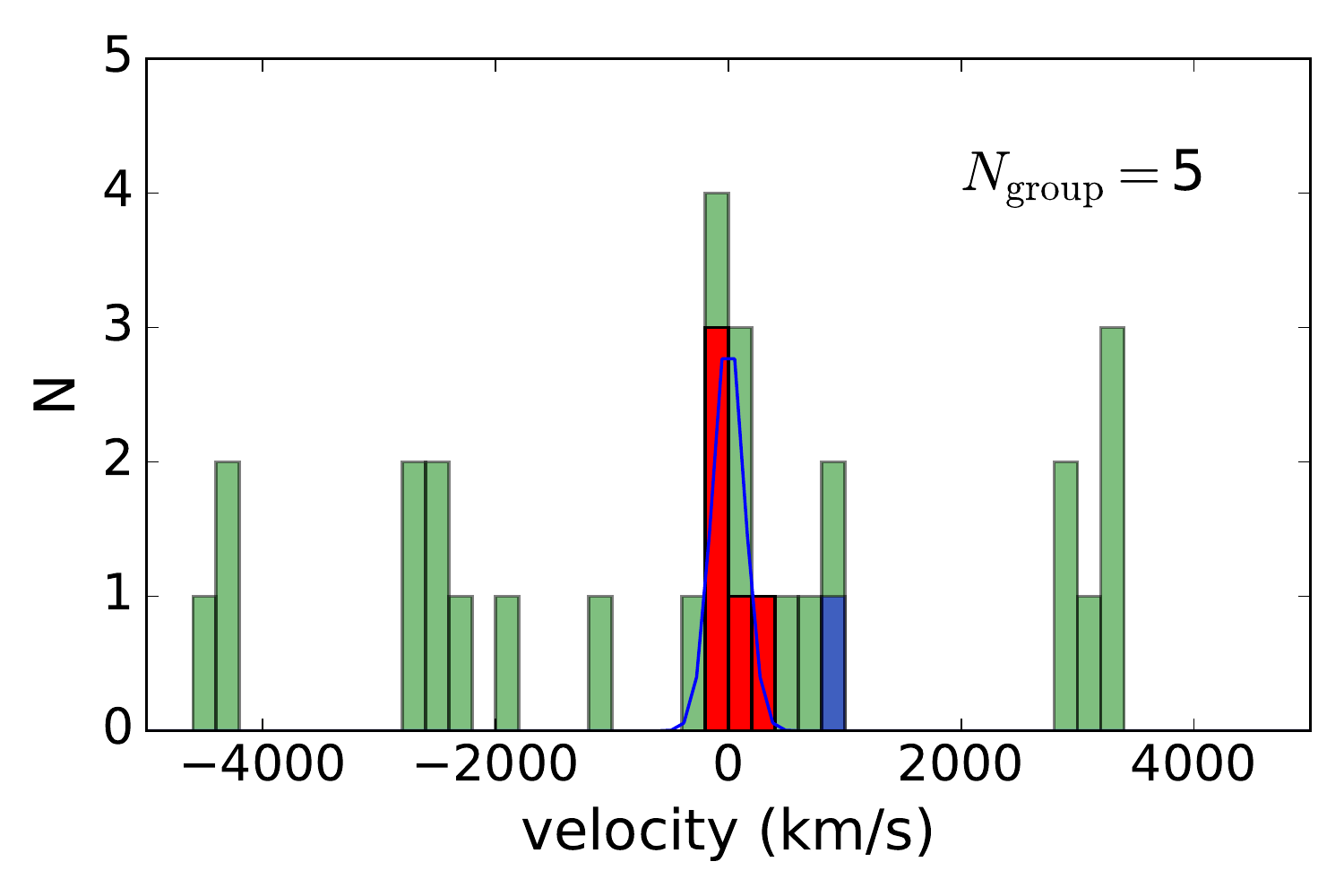}}
     & \subfloat{\includegraphics[scale=0.4]{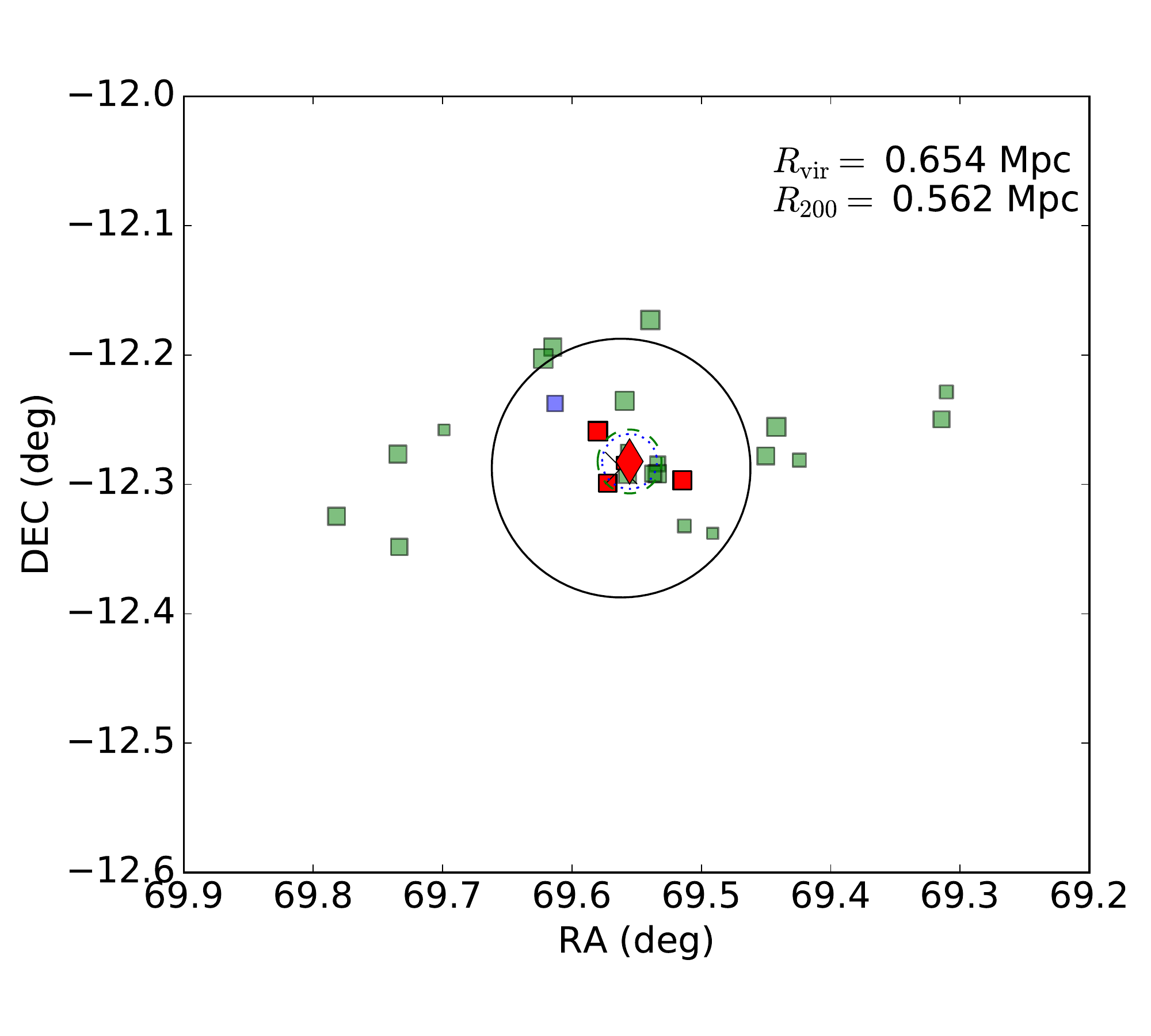}}	\\
     %	&	\subfloat{\includegraphics[scale=0.4]{HE0435_prop_mg8_360_m1.pdf}}  	
	%	\setcounter{subfigure}{9}

\end{tabular} 

	\setcounter{figure}{6}
	\caption{continued.}
\end{figure*}

\subsection{Groups: discussion}
\label{subsec:groupdiscuss}

For consistency, we can compare the number of groups we found with the average density of groups found in large surveys. A good comparison sample is the one from z-COSMOS \citep{Knobel2009} that identified spectroscopically (with 85\% completeness at $I_{AB} < 22.5\,$mag) 102 groups with $N \geq 5$ and $ 0.1 \leq z \leq 1$ in the 1.7 deg$^2$ COSMOS field. Rescaled to our field of view, an average number of $\sim 12$ groups would be expected, in good agreement with our results (i.e. 9 groups). Other works suggest a larger density of groups in that redshift range, but a direct comparison with our results is difficult due to the difference of selection techniques, magnitude limits, group definitions, group-richness densities \citep[e.g.][]{Milkeratis2010, Robotham2011, Gillis2011}.  \cref{After submission of this paper, \cite{Wilson2016} published a catalog of groups in the field of view of 28 galaxy-scale strong-lens systems based on the spectroscopic catalog of MOM15. The finding algorithm used by these authors is conceptually very similar to the one we present in Appendix~\ref{Appendix:methodII}, using the virial radius to set the group extent, but is based on a shallower spectroscopic catalog at small distance from the lens. The groups identified in the neighborhood of \HEofor\, agree between the two studies, with none of the groups identified by \cite{Wilson2016} missed by our algorithm. The group properties however sometimes differ, reflecting the dependance of group properties on the parameters used for group selection, and on the spectroscopic catalog. Three additional groups ($z=$0.1744, $z=$0.7019, $z=$0.4185) are found in our catalog, one of them ($z=$0.4185) at the same redshift as a visually idenfied group of $N=4$ galaxies reported by \cite{Wilson2016}. Because \cite{Wilson2016} initialize their group searches in different tiles around the lens, they more easily disentangle sub-groups where we report only visually identified bimodal group candidates. Overall, the two studies broadly agree and there is no evidence that our work is missing important structure towards \HEofor\, that would impact cosmological inference.} 

In the context of the cosmological inference from \HEofor\,(\HOLI Paper V), an important result from our search is the absence of very massive groups or galaxy clusters in the vicinity of the lens. However, several groups with a velocity dispersion $\sigma < 500$\,\kms\,are found. Five of them are found to lie within approximately one arcminute or less, from the lens. The richest of these groups is at the lens redshift, and has a velocity dispersion of about $\sigma=471\pm100$\,\kms. A similar group has been reported by \cite{Wong2011} in their analysis of the environment of nine strong lensing galaxies based on spectroscopy published by \cite{Momcheva2006}. More recently, \cite{Wilson2016} also studied the group properties of several lensed systems, and identified the same group as us at the lens redshift. The velocity dispersion of this group is similar in all three studies but the centroid differs by up to 50\arcsec\,due to our use of a luminosity weighted centroid\footnote{The group identified by \cite{Wong2011} does not include the galaxy (ID 6100) at ($\alpha, \delta$) = (69.439780, -12.223440). This shifts the centroid by $\sim$ 20\arcsec.}. 

The four other groups that appear in projection at separation of less than about one arcminute from the lens, are found at $\bar z_{\rm group} = 0.174$, $\bar z_{\rm group} = 0.419$, $\bar z_{\rm group} = 0.565$, $\bar z_{\rm group} = 0.702$. Although we do not identify any group at $z=0.78$, we could suspect the three galaxies G1, G2, G5 to be physically related with each other as they lie very close on the sky with a velocity spread of $\sim $ 360\,\kms. Another galaxy (ID 999) separated by less than 40\arcsec\, from these objects, could potentially be a member of the same group. Because we filter out tentative redshifts during the selection, and select groups only if $N > 4 $ galaxies, this system is not in our list of groups. If the group is only composed of G1-G2-G5, then our models including explicitly those galaxies (\HOLI Paper IV) should be sufficient to capture their perturbation of the gravitational potential. If other members were found (as potentially suggested by a small increase of galaxy counts with a $z_{\rm photometric} \sim 0.8$), the group centroid would likely move farther from the main lens and have a small impact on the lens model.

%-------------------------------------------------------------------------------

\section{Contribution of line of sight and environment to the lens structure}
\label{sec:model}

Modifications of the gravitational potential of the main lens produced by objects along the line of sight, or at the lens redshift, can be separated in two categories: i) perturbations that are weak enough to be approximated as a tidal perturbation (i.e. shear) and contribute as a constant external convergence to the main gravitational potential, and ii) perturbations that produce high order perturbations of the gravitational potential at the location of the lens (i.e. galaxies or galaxy groups yielding non negligible second and third order term in the Taylor expansion of the gravitational potential). In both cases, the amplitude of the effect depends on the redshift of the perturber. The strongest perturbations are caused by galaxies at the lens redshift or in the foreground of the main lens plane. Perturbers located behind the main lens need to appear closer in projection to the lens to yield high order perturbation of the potential \citep{McCully2016}. They can be otherwise approximated as a shear contribution, and their contribution to the convergence at the location of the lens be derived \citep{Fassnacht2006, Momcheva2006, Suyu2009, Suyu2013a, Greene2013, Collet2013}. In their work, \cite{McCully2014, McCully2016} have proposed a simple diagnostic to identify if a galaxy has to be treated explicitly in the lens model or if it can be accounted for as a tidal perturbation. For that purpose, one may compare the solutions of the lens equation in the tidal approximation when flexion produced by the perturber is included or not. For a point mass, the magnitude of the shift produced,  by the flexion term, called ``flexion shift'' $\Delta_3 x$, can be written:
\begin{equation}
\Delta_3 x = f(\beta) \, \times \frac{(\theta_{\rm E} \,\theta_{\rm E,p})^2}{\theta^3}, 
\end{equation}

\noindent where $\theta_{\rm E}$ and $\theta_{\rm E, p}$ are the Einstein radius of the main lens and of the perturber, and $\theta$ is the angular separation\footnote{This is the unlensed angular separation at the redshift of the perturber, which is almost equal to the observed one if the angular distance of the galaxy to the lens is sufficiently large compared to the lens angular Einstein radius.} on the sky between the lens and the perturber. The function $f(\beta) = (1-\beta)^2$ if the perturber is behind the main lens, and $f(\beta) = 1$ if the galaxy is in the foreground. In that expression, $\beta$ is the pre-factor of the lens deflection in the multiplane lens equation \citep[e.g.][]{SEF, Keeton2003a}. It encodes redshift differences in terms of distance ratios. For a galaxy at redshift $z_{\rm p} > z_{\rm d}$, we have:
\begin{equation}
\beta = \frac{D_{dp} D_{os}}{D_{op} D_{ds}},
\end{equation}

\noindent where the $D_{ij} = D(z_i, z_j)$ correspond to the angular diameter distance between redshift $z_i$ and $z_j$, and the subscripts o, d, p, s stand for the observer, deflector, perturber, and source. 

As long as the flexion shift of a galaxy is (much) smaller than the observational precision on the position of the lensed images, its perturbation on the gravitational potential of the main lens can be neglected in the lens model. \cite{McCully2016} shows, based on simulations and analysis of the line of sight of real lens galaxies, that perturbers with $\Delta_3 x >  10^{-4}$\, arcseconds need to be included explicitly in the modeling to avoid biasing $H_0$ at the percent level \citep[see e.g. Fig. 15 \& 16 of][]{McCully2016}. We should note that this cutoff is likely to be conservative as it is based on models constrained only by the quasar images fluxes, positions and time-delays, but not on the extended images of the host as performed in \HOLI. 

\subsection{Individual galaxies}

We first calculate the flexion shift for the individual galaxies in the field of \HEofor. For that purpose, we need to get a proxy on the Einstein radius $\theta_{\rm E, p}$ of those galaxies. First, we fix the redshift of the galaxies to their fiducial redshift in our spectroscopic catalog, if present, and to their photometric redshift otherwise. Second, we estimate the mass within $\theta_{\rm E, p}$ by rescaling the stellar mass derived in \HOLI Paper III to get a proxy on the total mass. For this purpose, we derive the dark matter contribution to the total mass within $\theta_{\rm E, p}$ (i.e. up to $\sim$ 60\% of the projected mass in the inner $\sim$\,10\,kpc of massive elliptical galaxies is dark matter) using the linear scaling relation between stellar mass and (projected) dark matter fraction in the Einstein radius derived by Auger et al. (\citeyear{Auger2010b}; Table 6). Galaxies with masses $M_* < 10^{10.5}\,M_{\odot}$ may have a larger contribution from their halo than what we would derive from extrapolating the relations from \cite{Auger2010b} to low mass \citep{Moster2010, Moster2013}, while not reducing drastically their Einstein radius due to their flatter inner mass density \citep{vandeven2009, Mandelbaum2009}. For those galaxies with $M_* < 10^{10.5}\,M_{\odot}$, we do not estimate the dark matter fraction, but use the scaling relation from \cite{Bernardi2011} to infer the velocity dispersion based on the stellar mass. We then assume that the galaxy can be modeled as a Singular Isothermal Sphere to derive its Einstein radius $\theta_{\rm E,p}$. 

\cref{In the above procedure, we use the most likely stellar mass from the photo-z catalog. This mass has been derived under the assumption of a Chabrier IMF, while there is evidence that IMF is not universal but more Salpeter-like at high mass \citep[e.g.][]{Barnabe2013, Posacki2015, Sonnenfeld2015a}. To account for this difference of IMF, we divide our stellar masses by a factor 0.55. Accordingly, we use the scaling relations from \cite{Auger2010b} that assume a constant Salpeter IMF. This choice of IMF has in practice almost no impact on the results since higher stellar masses for Salpeter IMF are compensated by lower dark matter fractions, yielding equivalent Einstein radii for the two IMFs.}    

\begin{figure}
% MXU_environment.figflexion()	
	\includegraphics[width=\linewidth]{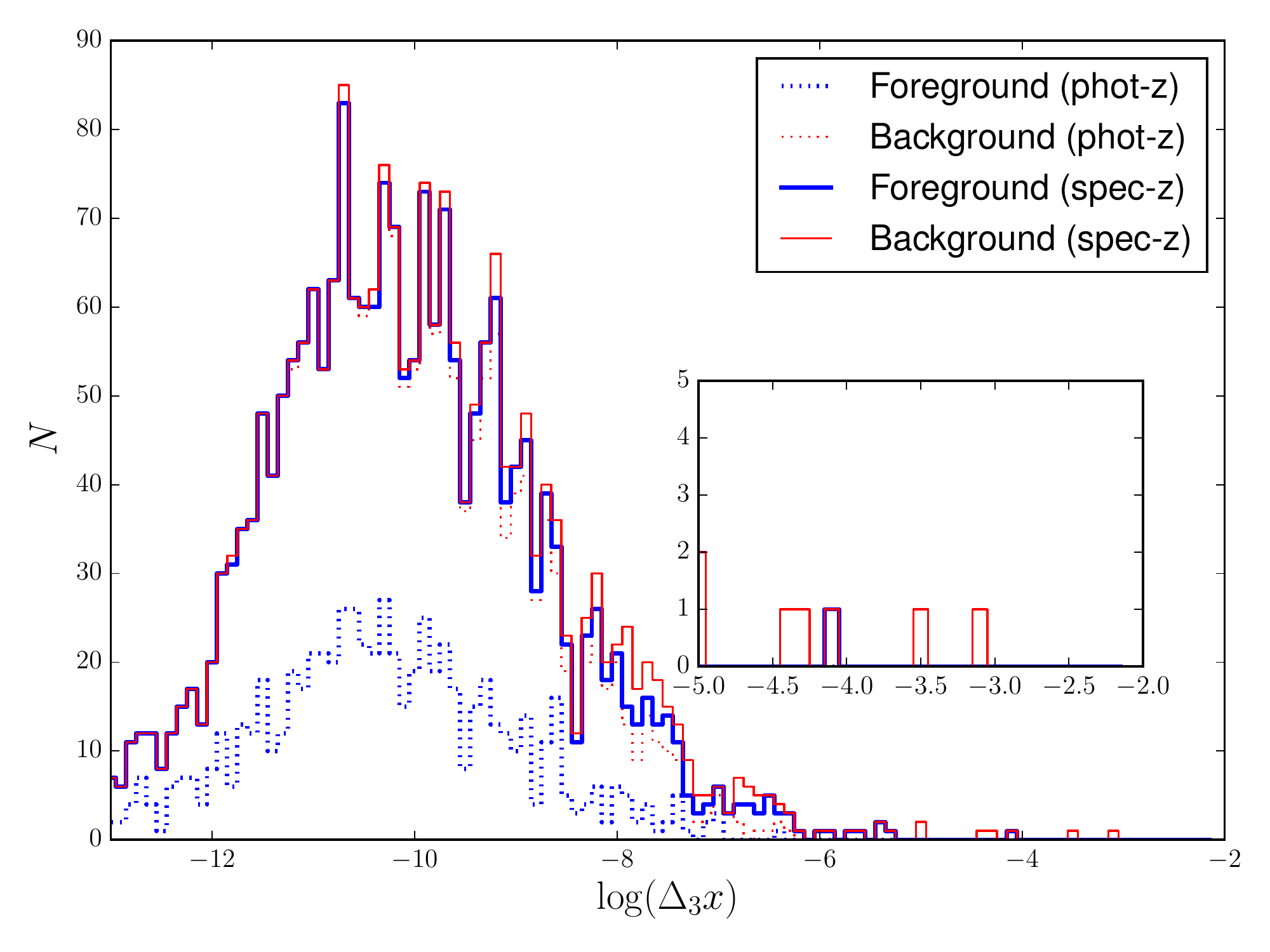}
	\caption{Distribution of maximum flexion shifts (in arcseconds; logarithmic scale) for the galaxies located within 6\arcmin\, of the lensing galaxy. The thick blue lines are for the galaxies at $z<z_{\rm d}$, while the thin red lines corresponds to galaxies with $z > z_{\rm d}$. Solid lines correspond to objects for which we have a spectroscopic redshift and dashed lines to galaxies for which we have only a photometric redshift. The inset panel displays a zoom of the region $10^{-5}$ arcsec $< \Delta_3 x < 10^{-2}$\,arcsec.  }
	\label{fig:flexion}
\end{figure}

Figure \ref{fig:flexion} shows the distribution of flexion shifts derived for all the galaxies located within 6\arcmin\,of \HEofor. \cref{The largest flexion shift is observed for three of the five galaxies closest in projection to \HEofor\,(i.e. G1, G3, G4, see Fig.~\ref{fig:Spec}) with $\Delta_3 x(\rm {G1, G3, G4})=(8.04\,\times 10^{-4}, 8.0\,\times 10^{-5}, 3.3\,\times 10^{-4} )$\,arcseconds. } The other galaxies have on their own little impact on main lens model. Despite that high order effects due to flexion combine in a complicated way \cref{(as the flexion shift is effectively a tensor)}, the sum of flexion shifts is interesting to calculate to verify that there is no subsample of galaxies that, together, would produce high order perturbations of the lens potential \citep{McCully2016}. The sum of flexions of all individual galaxies but G1-G3-G4, \cref{amounts $\sum_{i} \Delta_3 x_i \sim 1.6\,\times 10^{-4}\,$arcseconds,} providing a good indication that no (group of) additional objects need to be included explicitly in our models. \cref{This conclusion remains if we use the upper limit on the stellar mass to derive $\theta_{\rm E, p}$, as flexion shifts about 2-3 times larger are then derived.} In any case, $G1$ is the galaxy producing the largest perturbation of the lens potential, with a flexion shift several times larger than the other nearby galaxies. This motivates its explicit treatment in all the lens models presented in \HOLI Paper IV. Although we cannot rule out that the other galaxies play a role, their impact is substantially smaller. 

This very small perturbation of the environment and line of sight objects on the main gravitational potential of the lens is consistent with the number count analysis presented in \HOLI Paper III. This work demonstrates that the external convergence from the galaxies in the field of view of \HEofor\, combined with the existence of underdense lines of sight, yield a very small effective external convergence at the location of the lensed images. This is also in agreement with the weak lensing analysis of \HEofor\, (Tihhonova et al., in preparation) that finds a conservative $3\sigma$ upper limit of $\kappa_{\rm ext}=0.04$ at the lens position. 

\subsection{Flexion from groups}

Similarly to the approach followed for individual galaxies, we have calculated the flexion shift $\Delta_3 x$ associated to the groups. Because galaxies of a group host a common dark matter halo, they may have a larger impact on the lens potential than galaxies considered separately. We use the flexion shift to unveil if any of the identified groups has to be included explicitly in lens models.

Each group is described with a singular isothermal sphere model. Under this approximation, the Einstein radius $\theta_{\rm E, p}$ of a group at $\bar{z}_{\rm group}=z_{\rm p}$ is calculated from the distance ratios and intrinsic group velocity dispersion $\sigma_{\rm int}$:

\begin{equation}
\theta_{\rm E, p} = 4\pi \left(\frac{\sigma_{\rm int}}{c}\right)^2 \, \frac{D_{ps}}{D_{os}}.  
\end{equation}

In order to account for the uncertainty on the group centroid and velocity dispersions, we have estimated the flexion from 1000 bootstrap samples of these quantities. The distribution $\Delta_3 x$ derived from this technique follows roughly a log-normal distribution. Table~\ref{tab:groups} lists the value of $\log(\Delta_3 x)$ associated to the fiducial group and the standard deviation from the bootstrap distribution. We find that all the groups, except the group at the lens redshift, have negligible contribution to the flexion shift. In two cases, ($\bar{z}_{\rm group}=0.1744$ and $\bar{z}_{\rm group}=0.56372$), $\log(\Delta_3 x) > 10^{-4}$ arcseconds for up to 15\% of the bootstrap samples. Additional spectroscopic data may be needed to completely rule out a potential impact of those groups. 

The group hosting the lensing galaxy needs a separate discussion as $\Delta_3 x > 10^{-4}$\,arcseconds ($10^{-3}$\,arcseconds) for about 40\% (12\%) of the samples. In fact, this substantial chance for the group to impact the main lens potential is driven by the large uncertainty on the group centroid. However, we think that this uncertainty is overestimated by the bootstrapping approach as this technique assumes bootstrap samples with luminosities drawn from the observed luminosity of the group members. Since a luminosity weighting scheme is used to calculate the centroid, the bootstrapped centroids vary by much larger amount than if samples galaxy luminosities were drawn from the true underlying distribution of luminosity of the group members. Our analysis shows that most of the members of this group are separated by less than 3\arcmin\, from the lens, i.e. in a region where our spectroscopic completeness is the highest. Since we have not identified any new galaxy at the lens redshift in that region compared to MOM15\footnote{We have independently confirmed the redshift of four galaxies published in MOM15.}, we may consider that the group is complete down to $i=22\,$mag. We could therefore estimate the uncertainty on the group centroid by adding artificial fainter group members and re-estimating the centroid. Because of the luminosity weighting scheme, adding 10 galaxies with $i \in [22,24]\,$mag in a 3\arcmin\, radius field centered on the lens, shifts the centroid by typically 4\arcsec. This is not sufficient to increase the flexion shift above $10^{-4}$\,arcseconds. Alternatively, when considering a mass weighting scheme, we find a group position $\sim 67\arcsec$ away from the lens, but offset by 20\arcsec\, from the position reported in Table~\ref{tab:groups}. If we fix the group center to that position, we derive $\Delta_3 x \sim 7.9 \times 10^{-5}$ arcseconds, supporting a negligible role of the group on the lens model. 

The lens models presented in \HOLI Paper IV, when including only G1 or all the galaxies G1-G5 in the model, require additional external shear amplitude $\gamma_{\rm ext} \lesssim 0.03$. Such a small amount of shear from lens models is very often, but not systematically as the shear is a tensor, a good indication that perturbers are sufficiently distant to produce small changes of the lens potential \citep[see e.g.][]{Keeton1997, Holder2003, Sluse2012a}. If we model the group as an isothermal model (with $\theta_{\rm E, p} \sim 4\arcsec$, in agreement with the group properties in Table~\ref{tab:groups}), we find a shear $\gamma_{\rm group} \sim 0.035$ at the position of the lens. Similarly, assuming a circularly symmetric Navarro-Frenk-White profile \citep{Navarro1997}, with a concentration $c \equiv R_{\rm vir}/r_{s} = 5.1$ and a virial mass compatible with the virial mass reported in Appendix~\ref{appendix:Mvir} \citep{Verdugo2014, Viola2015}, we derive a shear amplitude $0.06 < \gamma_{\rm group} < 0.08$. The similar convergence $\kappa_{\rm group}$ expected from those models is difficult to reconcile with the 3$\sigma$ upper limit $\kappa_{\rm ext} < 0.04$ found in the weak lensing analysis of the field (Tihhonova et al., in preparation). This indicates that either the group centroid is even more distant from the lens than found through our luminosity weighting scheme, or that the lens lies at the center of its group halo as discussed hereafter. 

As the lens is the brightest group member, it is likely to be the center of its group halo \citep{Robotham2011, Shen2014a, Hoshino2015}. In that case, the lens models presented in \HOLI Paper IV would already account for the group halo. The projected dark matter fraction within the Einstein radius of the main lens is found, from the composite model (i.e. dark matter + baryons) presented in \HOLI Paper IV, to be $f_{DM} \sim 0.45$. This is in the range derived by SLACS \cite{Auger2010b} and SL2S \citep{Sonnenfeld2015a} for IMF between Chabrier and Salpeter. Considering the large intrinsic scatter in the fraction of dark matter within the Einstein radius of galaxies \citep[e.g.][]{Auger2010b, Xu2016}, this measurement is consistent with a modest excess of dark matter from the group halo in the lensing galaxy, as would be expected if the lens was at the group center.

%-------------------------------------------------------------------------------

\section{Conclusion}
\label{sec:conclude}

\cref {We have used multi-objects spectrographs on ESO-VLT, Keck and Gemini telescopes to measure the redshifts of 65 galaxies (down to $i = 23$\,mag) within a field of $\sim 4\arcmin$\, radius centered on \HEofor. In addition, our spectroscopic sample contains 18 galaxies with tentative redshifts, and 46 objects which had uncertain photometric classification, but turn out to be stars in our Galaxy. We have complemented our catalog with independent spectroscopic data sets compiled by MOM15. This expands the number of confirmed (or tentative) spectroscopic redshift in the field of \HEofor\, to 425 galaxies, up to a projected distance of 15\arcmin\, from the lens. Both the spectroscopic catalog and associated spectra are made publicly available with this paper.   }

The analysis of this new data set, combined with deep multicolor ($ugri$) photometric data covering the same field of view and presented in the companion \HOLI Paper III (Rusu et al., submitted), yields the following important results: 

\begin{enumerate}
\item The redshifts of the five brightest galaxies that fall within 12\arcsec\, of the lens (G1 - G5, with $i \in [19.9, 22.1]$\,mag), are measured to be $z_{\rm G1}=0.7821, z_{\rm G2}=0.7806, z_{\rm G3}=0.4191, z_{\rm G4}=0.4568, z_{\rm G5}=0.7792$, \cref{with a typical random uncertainty of $\sigma_z(\rm ran) \sim 0.0002$, and a possible systematic uncertainty $\sigma_z(\rm sys) \sim 0.0004$.}

\item In order to pinpoint the galaxies that are most likely to produce high order perturbations of the gravitational potential of the main lens, we have derived the so called flexion shift $\Delta_3 x$ \citep{McCully2016} of each individual galaxy in the field. \cite{McCully2016} suggest that $\Delta_3 x \sim 10^{-4}$ arcseconds is a conservative threshold above which a perturber is susceptible producing a bias at the percent level on $H_0$ if not included explicitly in the lens model. The largest flexion shift is found for G1 for which we get $\Delta_3 x(\rm {G1})\sim 8 \times 10^{-4}$\,arcseconds. This motivates the explicit inclusion of this galaxy in all the lens models of \HEofor\, presented in the companion \HOLI Paper IV (Wong et al., submitted). \cref{The two galaxies G3 and G4 are also found to have flexion shifts close to or above $10^{-4}$\,arcseconds} such that they are also included in one of the lens models presented in \HOLI Paper IV. 

\item We search for galaxy groups or clusters in the field of view of \HEofor\, using an iterative algorithm similar to those developed by \cite{Wilman2005}, \cite{Calvi2011} and \cite{Ammons2014}. Our iterative method identifies group members based on the joint separation of galaxies projected on the sky and redshift space. We have searched for galaxy groups of at least 5 members in the inner 6\arcmin\,around the lens, where our spectroscopic completeness is the highest, and for groups of at least 10 members at larger distance from the lens. No evidence for a massive galaxy cluster was found, but 9 galaxy groups (0.05 $< \bar{z}_{\rm group} < 0.8$) with velocity dispersion $\sigma_{int} < 500$\,\kms\,(some groups being possibly bimodal) were identified. One of these groups includes the lensing galaxy. It has been previously reported by \cite{Wong2011} with one less member, and is independently found by \cite{Wilson2016} based on the catalog published by MOM15. 

\item The impact of the groups on the lens model is more difficult to determine than for individual galaxies because of the uncertainty on the position of the group centroid. Fixing the group centroid to the brightest (spectroscopically confirmed) group member yields $\Delta_3 x < 10^{-4}$\,arcseconds for every group. A similar result is found when fixing the group centroid to the luminosity/mass weighted centroid of the identified group members. The centroid uncertainty has little impact on these conclusions for most of the groups but for the group hosting the lens. In that case, a shift of the luminosity/mass weighted centroid (found $\sim 70\arcsec$ from the lens), by more than 20\arcsec\, towards the lens would yield  a flexion shift a few times $10^{-4}$\,arcseconds. We think that such a shift is unlikely as we have good evidence that we identified all the members of that group down to $i\sim 22\,$ mag.

\end{enumerate}

Our spectroscopic study demonstrates that \HEofor\, requires an explicit inclusion of the nearest galaxy G1, while the galaxies G2-G5, produce smaller, but potentially non-negligible, perturbation of the gravitational potential of the main lens. On the other side, galaxy groups are unlikely to produce significant perturbations. This is confirmed by the weighted number counts analysis of the field of \HEofor\, presented in \HOLI Paper III, that shows that the line of sight is not particularly overdense, with an external convergence $\kappa_{\rm ext} = 0.003\pm 0.025$. The small convergence produced by the lens environment is confirmed by the weak lensing study of the field of view (Tihhanova et al., in prep) that shows that the total external convergence towards \HEofor\, is $\kappa_{\rm ext} < 0.04$ at $3\sigma$. This motivates the lens models presented in \HOLI Paper IV where only galaxy $G1$ is included explicitly in all the lens models using a mutiplane formalism, while a distribution of the convergence produced by the other galaxies (\HOLI Paper III), is used to account for the other galaxies. 

We are completing the analysis of the spectroscopic environment of the next two \HOLI  lensed systems, \HEeleven\, and \WFItwenty. The much richer line-of-sight environment of these two systems may produce stronger systematic errors on $H_0$ if not carefully accounted for in the lens models, making spectroscopic characterization of the lens environment a key ingredient of cosmography with time-delay lenses.  

%-------------------------------------------------------------------------------

\section*{Acknowledgments}

We thank Adriano Agnello, Roger Blandford, Geoff Chih-Fan Chen, Xuheng Ding,
Yashar Hezaveh, Kai Liao, John McKean, Danka Paraficz, Olga Tihhonova, and
Simona Vegetti for their contributions to the H0LiCOW project. H0LiCOW and COSMOGRAIL are made possible thanks to the continuous work
of all observers and technical staff obtaining the monitoring
observations, in particular at the Swiss Euler telescope at La Silla
Observatory. Euler is supported by the Swiss National Science
Foundation. 
We thank Michelle Wilson for sharing the lens galaxy group properties from an ongoing independent study. D.S. acknowledges funding support from a {\it {Back to Belgium}} grant from the Belgian Federal Science Policy (BELSPO), and partial funding from the Deutsche Forschungsgemeinschaft, reference SL172/1-1.
T.T. thanks the Packard Foundation for generous support through a Packard Research Fellowship, the NSF for funding through NSF grant AST-1450141, ``Collaborative Research: Accurate cosmology with strong gravitational lens time delays''.
\HOLI is supported by NASA through \emph{HST} grant HST-GO-12889.
S.H.S. acknowledges support from the Max Planck Society through the Max
Planck Research Group.  This work is supported in part by the Ministry
of Science and Technology in Taiwan via grant
MOST-103-2112-M-001-003-MY3.
K.C.W. is supported by an EACOA Fellowship awarded by the East Asia Core Observatories Association, which consists of the Academia Sinica Institute of Astronomy and Astrophysics, the National Astronomical Observatory of Japan, the National Astronomical Observatories of the Chinese Academy of Sciences, and the Korea Astronomy and Space Science Institute.
M.T. acknowledges support by a fellowship of the Alexander von Humboldt Foundation
V.B., F.C. and G.M. acknowledge the support of the Swiss National Science Foundation (SNSF).
S.H. acknowledges support by the DFG cluster of excellence \lq{}Origin and Structure of the Universe\rq{} (\href{http://www.universe-cluster.de}{\texttt{www.universe-cluster.de}}).
C.E.R and C.D.F. were funded through the NSF grant AST-1312329,
``Collaborative Research: Accurate cosmology with strong gravitational
lens time delays,'' and the HST grant GO-12889.
P.J.M. acknowledges support from the U.S.\ Department of Energy under
contract number DE-AC02-76SF00515.
LVEK is supported in part through an NWO-VICI career grant (project number 639.043.308).

% European Space Observatory 
Based on observations collected at the European Organisation for Astronomical Research in the Southern Hemisphere
under ESO programme(s) 091.A-0642(A) (PI: Sluse), and 074.A-0302(A) (PI:Rix). 
%Gemini
Based on observations obtained at the Gemini Observatory (GS-2013B-Q-28, PI: Treu), which is operated by the Association of Universities for Research in Astronomy, Inc., under a cooperative agreement with the NSF on behalf of the Gemini partnership: the National Science Foundation (United States), the National Research Council (Canada), CONICYT (Chile), Ministerio de Ciencia, Tecnolog\'{i}a e Innovaci\'{o}n Productiva (Argentina), and Minist\'{e}rio da Ci\^{e}ncia, Tecnologia e Inova\c{c}\~{a}o (Brazil). 
% Keck 
Some of the data presented herein were obtained at the W.M. Keck
Observatory, which is operated as a scientific partnership among the
California Institute of Technology, the University of California and
the National Aeronautics and Space Administration. The Observatory was
made possible by the generous financial support of the W.M. Keck
Foundation.  
The authors wish to recognize and acknowledge the very significant
cultural role and reverence that the summit of Mauna Kea has always
had within the indigenous Hawaiian community.  We are most fortunate
to have the opportunity to conduct observations from this mountain.
% Hubble Space Telescope
Based on observations made with the NASA/ESA Hubble Space Telescope, obtained at the Space Telescope Science Institute, which is operated by the Association of Universities for Research in Astronomy, Inc., under NASA contract NAS 5-26555. These observations are associated with program \#12889. Support for program \#12889 was provided by NASA through a grant from the Space Telescope Science Institute, which is operated by the Association of Universities for Research in Astronomy, Inc., under NASA contract NAS 5-26555. 
% Astropy and TOPCAT
This work made extensive use of TOPCAT \citep{Taylor2014}, and of Astropy, a community-developed core Python package for Astronomy \citep{Astropy2013}. \\

%%%%%%%%%%%%%%%%%%%%%%%%%%%%%%%%%%%%%%%%%%%%%%%%%%

%%%%%%%%%%%%%%%%%%%% REFERENCES %%%%%%%%%%%%%%%%%%

% The best way to enter references is to use BibTeX:
%\bibliographystyle{mnras}
%\bibliography{example} % if your bibtex file is called example.bib
% For github, build spec_environ.bib
\bibliography{spec_environ}

\bibliographystyle{mnras}
% Temporarily, I use
%\bibliography{/Users/sluse/work/articles/bibds}

%%%%%%%%%%%%%%%%%%%%%%%%%%%%%%%%%%%%%%%%%%%%%%%%%%

%%%%%%%%%%%%%%%%% APPENDICES %%%%%%%%%%%%%%%%%%%%%

\appendix

\section{Comparison with litterature redshift}
\label{Appendix:redshift}

\begin{figure}
	\includegraphics[width=\columnwidth]{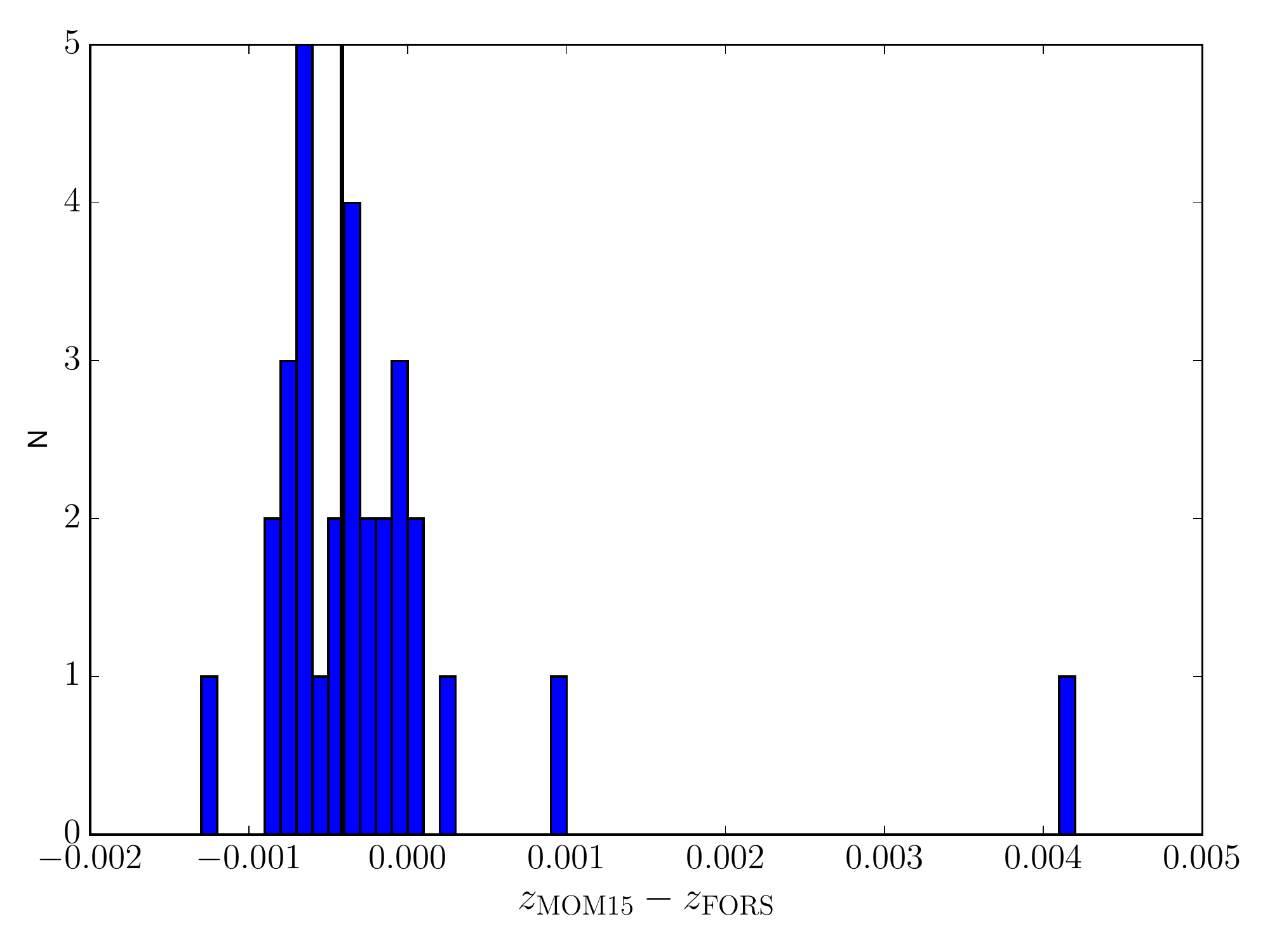}
		\caption{Distribution of the difference of redshifts  $\delta z = z_{\rm MOM15} - z_{\rm{FORS}}$ between measurements from MOM15 and our measurement on VLT-FORS spectra. The median redshift difference is shown with a black vertical line. }
	\label{fig:redshiftcompa}
\end{figure}

\cref{Figure~\ref{fig:redshiftcompa} shows the distribution of the difference of redshifts between MOM15 and our VLT-FORS measurements. The FORS redshifts are characterized by a median systematic offset $\delta z = z_{\rm MOM15} - z_{\rm{FORS}} = -0.0004$. This offset is likely caused by an un-identified systematic error in the wavelength calibration of the FORS data. This is supported by comparison performed for two other \HOLI lens systems (\HEeleven, \WFItwenty) for which we have obtained similar data but agreement with MOM15 (Sluse et al., in prep.). While the catalog is not corrected for this systematic error, a correction is applied for the group identification performed in Sect.~\ref{sec:groups}. Note that two additional galaxies measured by MOM15 and re-measured with our GMOS and Keck data, are not shown here as no information on systematic errors may be retrieved from so few measurements.}
% Add a note regarding galaxy 0059 that shows an offset much larger and in the other direction than MOM15 ? 

\section{Group identification based on $R_{\rm vir}$}
\label{Appendix:methodII}

A common alternative to the aspect ratio is to consider as part of a group those galaxies that are projected on the sky by less than a fraction of the group virial radius. This is by nature a difficult task as the virial radius depends on the characteristics of the group we are searching for. The method described in this section, uses the same group identification algorithm as the one presented in Sect.~\ref{subsec:algorithm}, but uses a criterion based on the virial radius to assess the group membership of a galaxy. 

As virial radius, we use $R_{200}$, the radius enclosing an overdensity of 200 with respect to the critical density, which can be estimated (Eq. 8 of Finn et al., \citeyear{Finn2005}) from the observed velocity dispersion of the group as: 

\begin{equation}
R_{200} = \frac{1.73 \, \sigma_{\rm obs}}{1000\,{\rm km\,s}^{-1}\,\sqrt{\Omega_\Lambda + \Omega_m\, (1+z)^3}} \, h^{-1} \, {\rm Mpc}. 
\label{eq:R200}
\end{equation}

As a cross check, we have also calculated the virial radius using the approximate formulation proposed by \cite[][ eq. 9]{Girardi1998}:
\begin{equation}
R_{\rm vir} = 0.002\,\sigma_{\rm obs},
\label{eq:Rvir}
\end{equation} 

\noindent where $\sigma_{\rm obs}$ is the observed velocity dispersion in units of \kms\,and $R_{\rm vir}$ is in Mpc. The two formula give consistent results to within 10\%. We therefore use only (\ref{eq:R200}) for group detection. 

Following \cite{Calvi2011}, we have used a 3 sigma clipping of the galaxies in redshift space (i.e. $n=3$ in (\ref{eq:dz})) and replace Eq.(\ref{eq:dtheta}) by: 
\begin{equation}
\delta \theta_{\rm max} = f_{\rm vir} \, R_{200} / D_\theta(z), 
\label{eq:dthetavir}
\end{equation}

\noindent with $f_{\rm vir} = 1.5$ \citep{Calvi2011}.

\cref {This equation expresses that the maximum angular transverse extension of the group is fixed to some fraction $f_{vir}$ of the angular virial radius. We may note that mostly two parameters influence the group detection, the clipping $n$ in redshift space (Eq.\ref{eq:dz}) and $f_{\rm vir}$. We experimented with $n=2$ and $f_{\rm vir} = 1$ and found those values to reduce the number and richness of detected groups, as expected as the new values effectively reduce the region of space where galaxy members are identified.}

\subsection{Results and discussion}

\cref{We present in Table~\ref{tab:groupsm3} the properties of the groups identified using $R_{\rm vir}$. The groups found using this method are similar to those presented in Table~\ref{tab:groups}, except the groups at $z=0.4185$ and $z=0.7019$ that are not found using this second method. The group properties sometimes differ between the two methods, especially for groups that are suspected to be bimodal, and groups identified based on a single FOV. }

% Group properties based on tab_mg/sg_360/900_m3.tex [Basically equivalent to what was called ``Method II'' in the first version of the paper (i.e. June 2016). Updated Table to fill the appendix (on Sept. 2016)
\begin{table*}
	% Based on tab_mg_360_m1.tex and tab_mg_360_m3.tex generated by MXU_environment.maingroups(rmax=360) 
	% Groups below do include a statistical correction of redshift
	\caption{\cref{Properties of the groups identified in the field of view of \HEofor, but using $R_{\rm vir}$ to select the groups (Appendix~\ref{Appendix:methodII}). The columns are the group redshift, the number of spectroscopically identified galaxies in the group, the group intrinsic velocity dispersion (rounded to 10\,\kms\, maximum precision) and 1$\sigma$ standard deviation from bootstrap, the group centroid, bootstrap error on the centroid, projected distance of the centroid to the lens, median flexion shift $\log(\Delta_3 x (arcsec))$ and $\sigma$ standard deviation from bootstrapping (Sect.~\ref{sec:model}). The last column indicate for which field of view the group is detected. The properties we display correspond to the field of view marked in bold.}}
	\label{tab:groupsm3}
	\centering
	\begin{minipage}{\linewidth}
		\centering
		\begin{tabular*}{0.8\linewidth}{lrcccccc}
			\hline
			$\bar{z}_{\rm group}$ & N & $\sigma_{int}$ $\pm$ err& $\alpha_{\rm ctr}$, $\delta_{\rm ctr}$ & err($\alpha_{\rm ctr}$, $\delta_{\rm ctr}$) &  $\Delta\theta$ & $\log(\Delta_3 x) \pm $ err & FOV  \\
			& & \kms & deg &  arcmin &  arcsec   & $\log(arcsec)$ & arcmin \\
						\hline
			0.0503& 10& 298 $\pm$121 & 69.627035, -12.325210 & 1.88, 1.47& 266.3& -6.28 $\pm$ 1.08 &  {\bf {15}} \\ %sg0 m3 , 900   V 
			0.1744& 6& 399 $\pm$ 82 & 69.548372, -12.280593 & 1.70, 2.00& 53.8& -5.25$\pm$ 1.36 & {\bf {6}} \\ %mg0, m3, 360    
			0.1853& 6& 605 $\pm$ 180 &69.619243, -12.310739 & 1.42, 1.13& 218.3& -5.14$\pm$ 1.09 &  {\bf {6}}  \\ % 360 mg1 m3  V
			0.31952& 10& 582$\pm$ 117 & 69.584638, -12.364008 & 1.39, 1.66& 287.1& -5.70$\pm$ 0.50 &  {\bf {15}} \\  % sg?? , 900, m3 
			0.4547& 11& 477$\pm$ 98 & 69.559414, -12.276026 & 0.36, 0.69& 41.9& -3.60$\pm$ 1.07 &  6, {\bf {15}} \\  % mg5 , 900arcsec, m3  
			0.5051$^\dagger$& 13& 441$\pm$ 95 & 69.581513, -12.233240 & 0.79, 0.45& 206.7& -5.99$\pm$ 0.57 &  6, {\bf {15}}  \\ % mg6, 900; m3 same with virial but 9 members. 
			0.56190$^\dagger$& 33& 1664$\pm$ 331 & 69.575620, -12.281520 & 1.11, 0.76& 52.4& -2.98$\pm$ 0.79 & 6, {\bf {15}} \\ %sg??, 360, m3 ;    different w. virial and different from 6arcmin -> bimodal

			\hline
		\end{tabular*}
		\begin{flushleft}
			{\small 
				{\bf Note:} $\dagger$ Possibly bimodal groups that may be constituted of 2 (or more) sub-groups. \\
			}
		\end{flushleft}
		
	\end{minipage}
	
\end{table*}

\section{Virial masses of the groups}
\label{appendix:Mvir}

An interesting physical property of the detected groups is their virial mass. The latter is however particularly difficult to estimate reliably \citep[see e.g.][]{Old2014}. Those masses are not used in our group selection process but may serve to verify the plausibility of a detected group. The virial theorem, applied to a stable system, yields a dynamical mass  $M \propto r\,\sigma^2$, where $r$ and $\sigma$ are the group radius and velocity dispersion. By further assuming that the group radius is proportional to the velocity dispersion \citep{Carlberg1997}, one finds that $M$ scales with $\sigma^3$. It is important to realize, that even if the virial theorem is well established, proxies to the group velocity dispersion (and radius) depends on the survey properties, such that the scaling relation depends also on the group selection technique and definitions choice of observational proxies to $r$ and $\sigma$ \citep{Old2014, Pearson2015}. To estimate the group masses, we use the relation\footnote{Since the relation from \cite{Pearson2015} was derived at $z < 0.1$, we folded in that relation the redshift dependence of $\sigma$, accounting for the fact that the velocity dispersion of a virialized system scales with $H(z)^{1/3}$.} $\log(M_{500}/(10^{14}\,M_\odot)) = \alpha\,\log((H_0/H(z))\times(\sigma/\sigma_0)^3) + \beta$, with ($\alpha, \beta, \sigma_0$)=(0.94, 0.39, 794.32 \kms) \citep{Pearson2015}. This relation, calibrated on X-ray mass $M_{500}$, and tested against systematics using mock data, shows rather large scatter and a systematic uncertainty of 0.3 dex, but has the advantage to be relatively robust against spectroscopic incompleteness \citep{Old2014, Pearson2015}. We derive $M_{200}$ from $M_{500}$ using $M_{200} = 1.38\,M_{500}$, which is exact for a NFW halo with concentration c=5 \citep{Newman2015}. Errors on $M_{200}$ are derived from error propagation on the scaling relation (i.e. accounting for the uncertainties on parameters $\alpha, \beta$ and on the measured $\sigma$). An uncertainty of 0.28 dex is quadratically added to the error on $\log(M_{200})$ to account for the systematic error derived from this relation by \cite{Pearson2015}. Table ~\ref{tab:Mvir} summarizes the virial mass and radius (derived from Eq.~\ref{eq:Rvir}) of the groups detected in this work. 

\begin{table}
	\caption{Virial mass, associated uncertainty, and radius of the groups identified in Sect.~\ref{sec:groups}.}
	\label{tab:Mvir}
	\begin{tabular}{lrr}
		\toprule
		$\bar{z}_{\rm group}$ & $\log(M_{\rm vir}/M_\odot) $ &   $R_{\rm vir} $ (Mpc)  \\
		\midrule
		0.0503 & 13.32$\pm$0.61 & 0.635  \\
		0.1744 & 13.81$\pm$0.40 & 1.071 \\
		0.1841 & 13.65$\pm$0.46 & 0.954  \\
		0.3202 & 13.83$\pm$0.36 & 1.259  \\
		0.4185 & 13.18$\pm$0.48 & 0.831  \\
		0.4547 & 13.72$\pm$0.36 & 1.385  \\
		0.5059 & 13.72$\pm$0.36 & 1.373 \\
		0.5650 & 13.33$\pm$0.43 & 1.062\\
		0.7019 & 12.49$\pm$0.63 & 0.654 \\
		\bottomrule
		
	\end{tabular}
\end{table}

%%%%%%%%%%%%%%%%%%%%%%%%%%%%%%%%%%%%%%%%%%%%%%%%%%

% Don't change these lines
\bsp	% typesetting comment
\label{lastpage}
\end{document}